\documentclass[useAMS,usenatbib]{mn2e}

\usepackage{amssymb,amsmath}
\usepackage{graphics}
\usepackage{subfigure}
\usepackage[dvips]{graphicx}
\usepackage{textcomp}
\usepackage{pdflscape}
\usepackage{epstopdf}
\usepackage{float}
\usepackage{xcolor}
\usepackage[percent]{overpic}
\def\ltsima{$\; \buildrel < \over \sim \;$}
\def\simlt{\lower.5ex\hbox{\ltsima}}   
\def\gtsima{$\; \buildrel > \over \sim \;$}
\def\simgt{\lower.5ex\hbox{\gtsima}}

\newcommand{\HI}{\textsc{H{\sc i}}}

\newcommand{\kms} {\,{\rm km\,s}^{-1}}
\newcommand{\cm} {\,{\rm cm}^{-3}}

\newcommand{\mo}{\,{\rm M}_\odot}

\newcommand{\gsim}{\lower.7ex\hbox{$\;\stackrel{\textstyle>}{\sim}\;$}}
\newcommand{\lsim}{\lower.7ex\hbox{$\;\stackrel{\textstyle<}{\sim}\;$}}
\newcommand{\s}{\sigma}
\newcommand{\sHa}{\sigma_{\textsc{H}_\alpha}}

\definecolor{brown}{rgb}{0.42,0.24,0.07}
\definecolor{darkgreen}{rgb}{0.0,0.6,0.00}
\definecolor{purple}{rgb}{0.7,0.0,0.7}
\definecolor{black}{rgb}{0.0,0.0,0.0}

\title[Modelling the SN-driven ISM]
{Modelling the supernova-driven ISM in different environments}

\author[A. Gatto, et al.]
{A. Gatto,$^{1}$\thanks{e-mail: andreag@mpa-garching.mpg.de}
S. Walch,$^{2,1}$ M.-M. Mac Low,$^{3}$ T. Naab,$^{1}$ P. Girichidis,$^{1}$ S. C. O. Glover,$^{4}$ \newauthor R. W\"{u}nsch,$^{5}$ R. S. Klessen,$^{4,6,7}$
P. C. Clark,$^{8,4}$ C. Baczynski,$^{4,3}$ T. Peters,$^{9,1}$ \newauthor
J. P. Ostriker,$^{10,11}$ J. C. Ib\'{a}\~{n}ez-Mej\'{\i}a$^{3,4}$ and S. Haid$^{2}$\\
$^{1}$Max-Planck-Institut f\"{u}r Astrophysik, Karl-Schwarzschild-Strasse 1, D-85748 Garching, Germany\\ 
$^{2}$Physikalisches Institut, Universit\"{a}t K\"{o}ln, Z\"{u}lpicher Strasse 77, D-50937 K\"{o}ln, Germany\\
$^{3}$Department of Astrophysics, American Museum of Natural History, 79th Street at Central Park West, New York, NY 10024, USA\\
$^{4}$Universit\"{a}t Heidelberg, Zentrum f\"{u}r Astronomie, Institut f\"{u}r Theoretische Astrophysik, Albert-Ueberle-Str. 2, D-69120 Heidelberg, Germany\\
$^{5}$Astronomick\'{y} \'{U}stav, Akademie v\u{e}d \u{C}eské Republiky, Bo\u{c}n\'{i} ́II 1401, C-14131 Praha, Czech Republic\\
$^{6}$Department of Astronomy and Astrophysics, University of California, 1156 High Street, Santa Cruz, CA 95064, USA\\
$^{7}$Kavli Institute for Particle Astrophysics and Cosmology, Stanford University, SLAC Nat. Acc. Lab., Menlo Park, CA 94025, USA\\
$^{8}$School of Physics and Astronomy, Cardiff University, 5 The Parade, Cardiff CF24 3AA, UK\\
$^{9}$Institut f\"{u}r Computergest\"{u}tzte Wissenschaften, Universit\"{a}t Z\"{u}rich, Winterthurerstrasse 190, CH-8057 Z\"{u}rich, Switzerland\\
$^{10}$Department of Astrophysical Sciences, Princeton University, 4 Ivy Ln, Princeton, NJ 08544, USA\\
$^{11}$Department of Astronomy, Columbia University, 550 West 120th Street, New York, NY 10027, USA
}

\begin{document}

\date{Accepted 2015 January 28. Received 2015 January 21; in original form 2014 October 31}

\pagerange{\pageref{firstpage}--\pageref{lastpage}} \pubyear{2015}

\maketitle

\label{firstpage}
%%%%%%%%%%%%%%%%%%%%%%%%%%%%%%%%%%%%%%%%%%%%

\begin{abstract}
We use hydrodynamical simulations in a $(256\;{\rm pc})^3$ periodic box to model the
impact of supernova (SN) explosions on the multi-phase interstellar medium (ISM) for
initial densities $n = $ 0.5--30~cm$^{-3}$ and SN rates 1--720~Myr$^{-1}$. We
include radiative cooling, diffuse heating, and the formation of molecular gas using
a chemical network. The SNe explode either at random positions, at density peaks, or
both. We further present a model combining thermal energy for resolved and momentum
input for unresolved SNe. Random driving at high SN rates results in hot gas
($T\gtrsim 10^6$ K) filling $> 90$\% of the volume. This gas reaches high
pressures ($10^4 < P/k_\mathrm{B} < 10^7$ K~cm$^{-3}$) due to the combination of 
SN explosions in the hot, low density medium and confinement in the periodic box.
These pressures move the gas from a two-phase equilibrium to the single-phase, cold branch of the cooling curve.
The molecular hydrogen dominates the mass ($>50$\%), residing in small, dense clumps.
Such a model might resemble the dense ISM in high-redshift galaxies.
Peak driving results in huge radiative losses, producing 
a filamentary ISM with virtually no hot gas, and a small
molecular hydrogen mass fraction ($\ll 1$\%).
Varying the ratio of peak to random SNe yields ISM
properties in between the two extremes, with a sharp transition for equal contributions.
The velocity dispersion in H{\sc i} remains $\lesssim 10 \kms$ in all cases. For peak driving
the velocity dispersion in H$_\alpha$ can be as high as $70 \kms$ due to the contribution from young, embedded SN remnants.
\end{abstract}

%%%%%%%%%%%%%%%%%%%%%%%%%%%%%%%%%%%%%%%%%%%%
%%%%%%%%%%%%%%%%%%%%%%%%%%%%%%%%%%%%%%%%%%%%

\section{ Introduction}\label{sec:introduction}
Supernova (SN) explosions are an important component for shaping the interstellar medium (ISM). In particular, they produce
its hottest phase, with temperature ${\rm T} \gtrsim 10^6\,{\rm K}$ \citep{CoxSmith74,McKeeOstriker77}. In the Milky Way, this hot phase fills from 20\% to 80\% of the volume with increasing height above the disc \citep{Ferriere01,KalberlaDedes08}.
SNe also drive turbulent motions in the ISM gas \citep{NormanFerrara96,JoungMacLow06,deAB07,Gent+13a}, 
although their relative importance with respect to other driving mechanisms has yet to be fully determined
\citep[see, e.g.,][]{MestelSpitzer56,SellwoodBalbus99,KritsukNorman02,Wada+02,Tamburro+09,KlessenHennebelle10}.

Supersonic turbulent motions in warm and cold gas, observed in emission from H{\sc i}, have been found in many local and extra-galactic environments, with a typical one-dimensional velocity dispersion of $\s \simeq 10 \kms$ \citep[e.g.][]{HeilesTroland03,PetricRupen07,Tamburro+09}. Similar velocity dispersions for molecular gas have been found from CO emission lines at galactic scales \citep{Caldu+13}, with a decrease to a few kilometers per second when observing small, isolated molecular structures \citep{Larson81} and a transition to transonic motions at the scale of dense cores \citep[$\sim 0.1\;{\rm pc}$;][]{Goodman+98}. Globally, these random motions could provide an effective turbulent pressure \citep{Ostriker+10} and help to regulate star formation \citep{MacLowKlessen04,ElmegreenScalo04}.

SNe are able to heat up the ISM, influencing its pressure and dispersing the gas locally \citep[e.g.][]{McKeeOstriker77,MacLow+05,JoungMacLow06,deAB04,deAB07,Joung+09}.
However, they could also sweep up the medium, causing colliding flows that may trigger new molecular cloud and star formation events \citep{ElmegreenLada77,Heitsch+06,VS+06,Banerjee+09,Heitsch+11}. 

Type {\sc ii} SN progenitors are massive OB stars with typically low space velocities of $\approx$ 15~km~s$^{-1}$ \citep{Stone91}. Due to these velocities, they can only travel tens to hundreds of parsecs away from their
birthplaces during their lifetimes.
However, 10--30\% of O stars and 5--10\% of B stars in the Galaxy are runaway stars, which have large velocities \citep[from 30 to a few hundred $\kms$;][]{Blaauw61,Gies87,GiesBolton96,Stone91} and are therefore found far from associations (from several hundred parsecs to a kiloparsec away from their birthplaces). Runaway stars are produced by dynamical ejection by means of gravitational scattering \citep[][]{GiesBolton96,FujiiZwart11,PeretsSubr12}, by massive binary systems where one of the two stars explodes as a SN, which leads to the ejection of the unbound companion 
\citep[][]{Blaauw61,Zwart00,Eldridge+11}, or both mechanisms \citep{AltenburgKroupa11}. 

Runaway massive stars typically explode in the diffuse ISM, i.e. in random positions, which are not correlated with the dense molecular clouds within which massive star formation takes place. This also applies to Type {\sc i}a SNe, which add an additional random component of SN explosions due to the long lifetimes of their progenitors.

A high fraction of massive stars (70--90 \%; \citealt{LadaLada03}) also form in clusters and associations, rather than in isolation.  
Despite being initially embedded in high density environments, the high degree of clustering and multiplicity of Type {\sc ii} SN progenitors causes the majority of SN explosions to happen within low density gas that has previously been processed by the combined effect of stellar winds \citep{TenorioTagle+90,TenorioTagle+91,BrighentiDercole94,RogersPittard13} and H{\sc ii} regions \citep{Gritschneder+09,Walch+12-2,Walch+13,Dale+14}, as well as earlier SNe in superbubbles \citep{McCrayKafatos87,MacLowMcCray88,TenorioTagleBodenheimer88}. However, there is evidence that SNe are also interacting with the dense gas. In the inner Galaxy, around 15\% of identified SN remnants show clear signs of interaction with molecular gas, including water maser emission \citep{HewittYusefZadeh09} and TeV $\gamma$-ray emission \citep{Fukui+03}.
Overall, SNe that explode in any of the above low density environments can be approximated by explosions at random positions in an ISM with a significant volume filling fraction (VFF) of hot gas. The remaining fraction of SNe interacting with dense gas can be modelled via explosions within the densest gas parcels \citep{WalchNaab14}.

Several studies of the SN-driven ISM have been carried out in the past, assuming clustered, random or density peak positions for the explosions. For instance, \citet[][]{deAB04} studied a representative piece of a stratified galactic disc, shaped by SNe going off at fixed rate and mostly placed in regions with highest density. \citet[][]{JoungMacLow06} performed similar simulations but chose random positions for their explosion locations.
In a recent paper, \cite{HennebelleIffrig14} perform magneto-hydrodynamical simulations of a stratified Galactic disc modelling star formation via star cluster sink particles. They show that SN explosions correlated in space and time with the particles' position and accretion can significantly inhibit further star formation. Similar conclusions have been drawn in \cite{IffrigHennebelle14}. A massive star exploding outside of the dense gas has a limited effect with respect to the case where the explosion is located within the cloud. In the latter case, a higher fraction of momentum is transferred to the cold gas and up to half of the cloud's mass can be removed by the SN.
Despite the number of previous works investigating the effect of SNe on both stratified \citep[e.g.][]{ShettyOstriker12, Hill+12, Gent+13a, KimOstriker+13} and unstratified ISM \citep[see, e.g.,][for modelling in periodic boxes]{Balsara+04,Kim04,Slyz+05,MacLow+05,Dib+06}, a systematic study is missing of the impact of different assumed SN positions, as well as the consequences of varying the SN rate with the gas density.

In this paper, we model the turbulent, multi-phase ISM in regions of different mean density over a timescale of $\sim100$ Myr with fixed (Type~{\sc ii}) SN rates, which are informed by the Kennicutt-Schmidt (KS) relation \citep{Schmidt59, Kennicutt98a}. We include a chemical network allowing us to partially model the effects of non-equilibrium chemistry in the warm gas, as well as the formation of H$_2$ and CO including dust and self-shielding.
We focus on three different scientific questions:
\begin{enumerate}
 \item Given a certain initial gas density, how do the properties of the ISM depend on the assumed SN rate?
 \item What are the differences between a medium shaped by SNe placed randomly or at density peaks?
 \item How does the ISM change as a function of the ratio between SNe that explode randomly or at density peaks?
\end{enumerate}

The paper is organised as follows: in section \ref{sec:method} we describe our model and runs.
In section \ref{sec:results} we show our results. Finally,
we present our conclusions in section \ref{sec:conclusions}.

%%%%%%%%%%%%%%%%%%%%%%%%%%%%%%%%%%%%%%%%%%%%
%%%%%%%%%%%%%%%%%%%%%%%%%%%%%%%%%%%%%%%%%%%%

\section{Numerical method, parameters and runs}\label{sec:method}
\subsection{Simulation method}
We use the Eulerian, adaptive mesh refinement, hydrodynamic code \textsc{FLASH} 4 \citep{flash,Dubey+08,Dubey+13} to model the SN-driven ISM in three-dimensional (3D) simulations. We use the directionally split, Bouchut HLL5R solver \citep{Bouchut+07,Waagan09,Bouchut+10} to simulate a volume of ($256$~pc)$^3$ with periodic boundary conditions. For the majority of the simulations the resolution is fixed to 128$^3$ cells ($\Delta x = 2$ pc), but we run a few setups at higher resolution (see section \ref{sec:higres}).
In different simulations, we change the initial total number density of the box from $n_\mathrm{i} =0.5\;{\rm cm}^{-3}$ to $n_\mathrm{i} =30\;{\rm cm}^{-3}$ (see Table \ref{tab:nSNR}) and the initial temperature is 6000 K.
%%%%%%%%%%%%%%%%%%%%%%%%%%%%%%%%%%%%%%%%%%%%

\subsection{Gas cooling and chemistry}
We include a chemical network to treat radiative cooling and diffuse heating as well as molecule formation. In the cold and warm gas, the main contributions to the cooling rate come from Lyman-$\alpha$ cooling, H$_{2}$ ro-vibrational line cooling, fine structure emission from C{\sc ii} and O, and rotational line emission from CO. These processes are modelled using the prescription developed in \cite{GloverMacLow07a,GloverMacLow07b}, \cite{Glover+10} and \cite{GloverClark12}. The chemical network therefore follows the abundances of five key chemical species: H{\sc i}, H{\sc ii}, H$_{2}$, C{\sc ii} and CO. 
For the formation of molecular hydrogen and CO, the effects of dust shielding and molecular (self-)shielding are included using the \textsc{TreeCol} algorithm of \cite{Clark+12}. Full details of the implementation of these processes  
within \textsc{FLASH} 4 are described in \cite{SILCC-1} and W\"{u}nsch et al.~(in prep.).

In hot gas, the excitation of helium and of partially ionised metals (e.g.\ O{\sc vi}) becomes important. The contribution to the cooling rate from these processes is computed from the cooling rates of \cite{GnatFerland12}, which assume that the atoms and ions are in collisional ionisation equilibrium. 
However, we explicitly follow the hydrogen ionisation state since non-equilibrium effects are noticeable for gas at temperatures around $10^{4}$~K \citep[see, e.g.,][]{Walch+11,Micic+13}. 

Diffuse heating from the photoelectric effect, cosmic rays, and X-rays is included following the prescriptions of \cite{Glover+10} and \cite{GloverClark12}. We assume a constant radiation field where the photo-ionisation from point sources is neglected. The far-UV interstellar radiation field is $G_{\mathrm 0} = 1.7$ \citep{Habing68,Draine78}, while the cosmic ray ionisation rate of H{\sc i} is $\zeta = 3 \times 10^{-17}$ s$^{-1}$. For the X-ray ionisation and heating rates we use the values of \cite{Wolfire+95}. The (constant) dust-to-gas mass ratio is set to $10^{-2}$.
We assume standard solar ratio of hydrogen to helium and solar metallicity with abundances $x_{\mathrm{O, Si}} =3.16 \times 10^{-4}$, $x_{\mathrm{Si\textsc{ii}}} = 1.5 \times 10^{-5}$ \citep{Sembach+00}, and $x_{\mathrm{C, tot}} = 1.41 \times 10^{-4}$. At $t=0$ the gas has an ionisation fraction of $0.1$.

%%%%%%%%%%%%%%%%%%%%%%%%%%%%%%%%%%%%%%%%%%%%

\subsection{Initial turbulent stirring}\label{sec:stirring}
At first, the gas is stirred with an Ornstein-Uhlenbeck random process \citep{EswaranPope88} with dimensionless wavenumbers
$k = 1$--2, where $k = 1$ corresponds to the box side, $L$. The phase turnover time is $t_\mathrm{drive}=25$ Myr, which corresponds to one crossing time for the warm gas. We distribute the turbulent energy onto solenoidal and compressive modes, using a 2:1 ratio. At every time step the total energy input is adjusted such that we obtain a global, mass-weighted, 3D root-mean-square (rms) velocity that is roughly constant at $v_\mathrm{3D, rms}=10 \kms$. 
We stir the gas for one crossing time (25 Myr) in order to generate a turbulent two-phase medium before the onset of SN driving \citep[see also][]{Walch+11}.
%%%%%%%%%%%%%%%%%%%%%%%%%%%%%%%%%%%%%%%%%%%%

\subsection{Supernova driving}\label{sec:SNdriving}
After 25 Myr, the artificial turbulent driving is stopped. Instead, we initialise SN explosions at a fixed rate, which we adjust to the mean box mass density $\rho_\mathrm{i}$. We compute the gas surface density of the box as $\Sigma_{\rm gas} = \rho_\mathrm{i} L$. The Kennicutt-Schmidt relation \citep{Kennicutt98a}:
\begin{equation} \label{eq:KS}
\frac{\Sigma_{\mathrm{SFR}}}{\mo \mathrm{yr}^{-1} \mathrm{kpc}^{-2}}=2.5 \times 10^{-4}\ \Bigl(\frac{\Sigma_{\rm gas}}{\mo  \mathrm{pc}^{-2}} \Bigr)^{1.4}\ ,
\end{equation}
relates the total gas surface density $\Sigma_{\rm gas} = \Sigma_{\mathrm{H\textsc{i}+H_2}}$ to a typical star formation rate (SFR) surface density. For a conventional stellar initial mass function \citep[e.g.][]{Salpeter55}, approximately 100 $\mo$ of gas that collapses into stars produces on average only one massive star, which will explode as a SN Type {\sc ii} at the end of its lifetime. Hence, we only consider Type {\sc ii} SNe and neglect the additional contribution from Type {\sc i}a SNe, which is $\sim15$\% of the total SN rate in the Galaxy \citep{Tammann+94}. With this information we compute the SN rate, $\dot{N}_{_{\rm SN, KS}}$, for every $n_\mathrm{i}$ as
\begin{equation}\label{eq:SNR}
\frac{\dot{N}_{_{\rm SN, KS}}}{{\rm Myr}} = \frac{\Sigma_{\mathrm{SFR}}}{\mo \mathrm{Myr}^{-1} \mathrm{pc}^{-2}} \times 10^{-2}  \frac{L^2}{\mathrm{pc}^{2}}\ .
\end{equation}
The SN rate derived in this way is afflicted with uncertainties of at least a factor of 2. For this reason we vary the SN rate by a factor of 2 for a number of initial densities. We list $\dot{N}_{_{\rm SN}}$ for each density in Table \ref{tab:nSNR}.

%%%%%%%%%%%%%%%%%%%%%%%%%%%%%%%%%%%%%%%%%%%%
%%%%%%%%%%%%%%%%%%%%%%%%%%%%%%%%%%%%%%%%%%%%

\begin{table}
\centering
\begin{tabular}{c|c|c|c|c}
\hline
    $n_\mathrm{i}$       &$\Sigma_{\rm gas}$ & $\dot{N}_{_{\rm SN, -}}$ & $\dot{N}_{_{\rm SN, KS}}$ & $\dot{N}_{_{\rm SN, +}}$ \\
(cm$^{-3}$) & (M$_\odot\ \mathrm{pc}^{-2}$)    & (Myr$^{-1}$)     & (Myr$^{-1}$)          & (Myr$^{-1}$)\\
\hline
0.5         &4.1  &-    &1.2  &-\\
1           &8.1  &1.5  &3    &6\\
3           &24.3 &7    &14   &28\\
10          &81.2 &38.5 &77   &154\\
30          &243  &180  &360  &720\\
\hline
\end{tabular}
\caption{Parameters of the simulations. Column 1 gives the mean volume density $n_\mathrm{i}$ and column 2 gives the corresponding gas surface density. In column 3 to 5 we list the simulated SN rates, where $\dot{N}_{_{\rm SN, KS}}$ is derived from eq. (\ref{eq:SNR}).}
\label{tab:nSNR}
\end{table}
%%%%%%%%%%%%%%%%%%%%%%%%%%%%%%%%%%%%%%%%%%%%
%%%%%%%%%%%%%%%%%%%%%%%%%%%%%%%%%%%%%%%%%%%%

\subsection{Supernova energy and momentum input}\label{sec:SN}
For the majority of our simulations, each SN injects $E_{\mathrm{SN}}=10^{51}$ erg of thermal energy into the ISM. 
We distribute this energy within a sphere with radius $R_\mathrm{inj}$. The radius of the injection region is adjusted such that it encloses $10^3\,\mo$ of gas, but we require $R_\mathrm{inj}$ to be resolved with a minimum of 4 cells:
\begin{equation} \label{Rinj}
R_\mathrm{inj} = \begin{cases}
\Bigl(\dfrac{3}{4 \mathrm{\pi}} \dfrac{10^3 \mathrm{M}_\odot}{\overline{\rho}}\Bigr)^\frac{1}{3} &\text{if $R_\mathrm{inj} \geqslant 4 \Delta x$}\\
4 \Delta x &\text{if $R_\mathrm{inj} < 4 \Delta x$}\ ,
\end{cases}
\end{equation}
where $\overline{\rho}$ is the mean density within the injection region.
Here, we present a SN model that combines the injection of thermal energy and momentum and which adjusts to the local environment of each SN explosion (see section \ref{sec:momentum}).

\subsubsection{ Thermal energy input}
If the SNe are resolved, we inject all of the SN energy in the form of thermal energy. Typically, the gas within $R_\mathrm{inj}$ is then heated up to $\approx 10^6$--$10^7$ K, which corresponds to a local sound speed $c_\mathrm{s}$ of a few hundred ${\rm km\ s}^{-1}$. 
The associated pressure increase causes a Sedov-Taylor blast wave to expand into the ambient ISM.
We decrease the time-step according to a modified Courant-Friedrichs-Lewy (CFL) condition to capture the dynamics of the blast wave:
\begin{equation} 
\Delta t = C_{\mathrm{CFL}} \frac{\Delta x}{\mathrm{max}(|v|+c_\mathrm{s})} \ .
\end{equation}

\subsubsection{ Momentum input}\label{combined_model}
In case of high densities within the injection region, the Sedov-Taylor phase of the SN remnant is unresolved. The mass within the injection region is high ($M_\mathrm{inj}>$ few $\times\ 10^3 \mo$), and therefore the thermal energy input would result in an effective temperature below $10^6$ K:
\begin{equation} \label{TiSN}
T_{\mathrm{i, SN}}= (\gamma - 1) \dfrac{E_{\mathrm{SN}}}{M_\mathrm{inj}} \dfrac{\mu m_\mathrm{p}}{k_\mathrm{B}}\ < 10^6 \mbox{ K}\ ,
\end{equation}
where $\gamma$ is the polytropic index, $\mu$ the mean molecular weight, $m_\mathrm{p}$ the proton mass and $k_\mathrm{B}$ the Boltzmann constant.

The cooling rate is a non-linear function of the gas temperature. At $T<10^6$ K the injected energy is almost instantaneously lost due to strong radiative cooling. The employment of a limited mass and/or spatial resolution may then lead to over-cooling, which is a well-known problem in galaxy simulations \citep[e.g.][]{Stinson+06, Creasey+11, Gatto+13}. Many different solutions have been put forward to address the problem of unresolved SNe: switching off cooling for a certain time after the explosion \citep[e.g.][]{ThackerCouchman00}; clustering of massive stars to develop superbubbles \citep[e.g.][]{ShullSaken95,Krause+13,Keller+14,Sharma+14};
or momentum rather than energy input \citep[e.g.][]{Kim+11,ShettyOstriker12}. Of course one could just increase the resolution (see section \ref{sec:higres}).

We choose a momentum input scheme for unresolved SNe, which is based on \citet{Blondin+98}.
For each explosion, we first calculate the radius of the bubble at the end of the Sedov-Taylor phase \citep{Blondin+98}
\begin{equation} \label{RST}
R_{\mathrm{ST}} = 19.1\ \Bigl(\dfrac{E_{\mathrm{SN}}}{10^{51}\
  \mathrm{erg}}\Bigr)^\frac{5}{17}\ \Bigl(\dfrac{\overline{n}}{\cm}\Bigr)^{-\frac{7}{17}}\ 
\mathrm{pc},
\end{equation}
where $\overline{n}$ is the mean number density within the injection region. If $R_{\mathrm{ST}} < 4\Delta x$, then we inject momentum rather than thermal energy \citep[see also][]{Hopkins+14}. The momentum is computed from \citep{Blondin+98}
\begin{equation} \label{PST}
p_{\mathrm{ST}} = 2.6\times 10^5 \Bigl(\dfrac{E_{\mathrm{SN}}}{10^{51}\ \mathrm{erg}}\Bigr)^\frac{16}{17}\Bigl(\dfrac{\overline{n}}{\cm}\Bigr)^{-\frac{2}{17}}
\mo\ \mathrm{km\ s^{-1}}.
\end{equation}
We deposit this momentum to the flattened density distribution within the injection region by adding the corresponding velocity of 
\begin{equation} \label{vinj}
v_{\mathrm{inj}} = \dfrac{p_{\mathrm{ST}}}{M_{\mathrm{inj}}}\ ,
\end{equation}
where $v_{\mathrm{inj}}$ points radially outwards. In addition, we increase the temperature of the injection region to 10$^4$ K.
This guarantees that, despite the significant energy losses, the momentum input is accounted for.

This method has two caveats. 
i) The momentum of a SN bubble can still increase during the pressure-driven snowplough phase \citep{Cioffi+88}, which is not included in this model. 
ii) Due to the injection of transonic or subsonic motions, high Mach number shocks are not created and little
to no hot gas is produced. 
Therefore, the momentum input given in equation (\ref{PST}) is a lower limit.

Recent results by \citet[][see also \citealt{Martizzi+14,Simpson+14}]{KimOstriker14} indicate that the final momentum driven by a single SN explosion can be captured (within 25\% of the expected value from high resolution, sub-pc simulations) under the conditions: (i) $R_\mathrm{ST} >3\ R_\mathrm{inj}$ and (ii) $R_\mathrm{ST} > 3\ \Delta x$. This is roughly insensitive to the ambient medium density distribution. As a result, the impact of SNe exploding in density peaks
might be underestimated even when $T_{\mathrm{i, SN}} > 10^6$ K for thermal energy injection.
In section \ref{sec:higres} we compare these requirements with our results at different resolutions.

\subsubsection{Supernova positions and simulations}\label{sec:runs}
%%%%%%%%%%%%%%%%%%%%%%%%%%%%%%%%%%%%%%%%%%%%
%%%%%%%%%%%%%%%%%%%%%%%%%%%%%%%%%%%%%%%%%%%%
\begin{table}
\centering
\begin{tabular}{ccccc}
\hline
name    		&$n_\mathrm{i}$       & $\dot{N}_{_{\rm SN}}$              & Driving   & Note \\
			&(cm$^{-3}$) & (Myr$^{-1}$)     &           &      \\
\hline
R-n$_{0.5}$         	&0.5  	     &1.2               &R          &\\
R-n$_\mathrm{1}$ 	&1  	     &3                 &R          &\\
R-n$_{1-}$ 		&1  	     &1.5               &R          &\\
R-n$_{1+}$ 		&1  	     &6                 &R          &\\
R-n$_\mathrm{3}$	&3  	     &14                &R          &\\
R-n$_{3-}$		&3  	     &7                 &R          &\\
R-n$_{3+}$		&3  	     &28                &R          &\\
R-n$_\mathrm{10}$	&10  	     &77                &R          &\\
R-n$_{10-}$		&10  	     &38.5              &R          &\\
R-n$_{10+}$		&10  	     &154               &R          &\\
R-n$_\mathrm{30}$	&30  	     &360               &R          &\\
R-n$_{30-}$		&30  	     &180               &R          &\\
R-n$_{30+}$		&30  	     &720               &R          &\\
\hline
P-n$_{0.5}$         	&0.5  	     &1.2               &P          &\\
P-n$_\mathrm{1}$ 	&1  	     &3                 &P          &\\
P-n$_\mathrm{3}$	&3  	     &14                &P          &\\
P-n$_\mathrm{10}$	&10  	     &77                &P          &\\
\hline
M10-n$_\mathrm{3}$	&3  	     &14                &M          &$f_\mathrm{peak}=10$\\
M20-n$_\mathrm{3}$	&3  	     &14                &M          &$f_\mathrm{peak}=20$\\
M30-n$_\mathrm{3}$	&3  	     &14                &M          &$f_\mathrm{peak}=30$\\
M40-n$_\mathrm{3}$	&3  	     &14                &M          &$f_\mathrm{peak}=40$\\
M50-n$_\mathrm{3}$	&3  	     &14                &M          &$f_\mathrm{peak}=50$\\
M60-n$_\mathrm{3}$	&3  	     &14                &M          &$f_\mathrm{peak}=60$\\
M70-n$_\mathrm{3}$	&3  	     &14                &M          &$f_\mathrm{peak}=70$\\
M80-n$_\mathrm{3}$	&3  	     &14                &M          &$f_\mathrm{peak}=80$\\
M90-n$_\mathrm{3}$	&3  	     &14                &M          &$f_\mathrm{peak}=90$\\
\hline
P-C-n$_\mathrm{3}$	&3  	     &14                &P          &CM\\
M50-C-n$_\mathrm{3}$	&3  	     &14                &M          &CM $f_\mathrm{peak}=50$\\
R-HR-n$_\mathrm{3}$	&3  	     &14                &R          &$\Delta x = 1$ pc\\
P-HR-n$_\mathrm{3}$	&3  	     &14                &P          &$\Delta x = 1$ pc\\
\hline
\end{tabular}
\caption{List of all simulations. From left to right: name of the simulation;
initial number density of the box $n_\mathrm{i}$; supernova rate $\dot{N}_{_{\rm SN}}$; Driving mode: pure random (R), pure peak (P) or mixed (M); Note: combined model with thermal energy and momentum input (CM).}
\label{tab:runs}
\end{table}
%%%%%%%%%%%%%%%%%%%%%%%%%%%%%%%%%%%%%%%%%%%%
We will show that the positioning of the SNe relative to the gas density distribution changes the structure of the ISM. Therefore, we choose three different schemes to place the SNe: (i) random driving (runs R-n$_\mathrm{i}$), (ii) peak driving (runs P-n$_\mathrm{i}$), i.e. the current SN is placed on the global density maximum at the time, and (iii) a mixture of the two, i.e. mixed driving (runs Mf-n$_\mathrm{i}$), with a fixed ratio $f_\mathrm{peak}$ of peak to total SNe.

We carry out a number of simulations with different driving schemes and densities, which we summarise in Table \ref{tab:runs}.
By default, we choose random driving. For this driving scheme, we run five simulations with $\dot{N}_{_{\rm SN, KS}}$ for $n_\mathrm{i} = 0.5, 1, 3, 10, 30 \cm$ and eight for $n_\mathrm{i} = 1, 3, 10, 30 \cm$ with half ($\dot{N}_{_{\rm SN, -}}$) and twice the KS SN rate  ($\dot{N}_{_{\rm SN, +}}$). In addition, we perform four simulations ($n_\mathrm{i} = 0.5, 1, 3, 10 \cm$) with peak driving. For $n_\mathrm{i}=3 \cm$, we carry out 9 simulations using mixed driving with different values of $f_\mathrm{peak}$.

We investigate the applicability of the combined model of thermal energy and momentum injection using two simulations with $n_\mathrm{i}=3 \cm$ and $f_\mathrm{peak}=100$\% (P-C-n$_\mathrm{3}$) and 50\% (M50-C-n$_\mathrm{3}$).
Finally, we run two simulations for $n_\mathrm{i}=3 \cm$ at a higher resolution (with $\Delta x=1$ pc).
These runs are called R-HR-n$_\mathrm{3}$ for random and P-HR-n$_\mathrm{3}$ for peak driving. 

\subsection{Simulation analysis}\label{sec:sims}
The simulations are stopped once an approximate chemo-dynamical equilibrium is reached, i.e. the mass distribution of the different chemical species stays roughly constant. This is typically the case after 100--200 Myr (note that 25 Myr was the initial crossing time), depending on the initial box density and SN rate.
We discuss the properties of the resulting ISM for different initial densities, driving schemes, and resolution towards the end of the simulations.
In particular, we focus on four quantities:
\begin{enumerate}
\item the distribution of gas mass among ionised, neutral, and molecular hydrogen;
\item the gas pressure in different temperature regimes; 
\item the VFFs of gas within the different temperature regimes; and 
\item the observable velocity dispersion. 
\end{enumerate}
We compute every quantity by averaging over the final 5~Myr of each simulation in order to reduce statistical errors. 
In Appendix \ref{sec:appendix} we show the full time history for three simulations with $n_\mathrm{i}=3 \cm$ using random and peak driving.

To compare with velocity dispersions observed in H{\sc i} and H$_\alpha$, we compute the mass-weighted, one-dimensional velocity dispersion, $\sigma_\mathrm{avg \;1D}$. We first compute the 1D dispersion for H{\sc i} and H$_\alpha$ along the x-, y-, and z-directions for each simulation:
\begin{equation} \label{sigmaturb}
\sigma_{j,k, {\rm turb}}=\left(\dfrac{\sum_i (v_{i,k}-\overline{v}_k)^2\ m_{i,j}}{M_{j,\rm{tot}}}\right)^{1/2}\ ,
\end{equation}
where $j$ indicates the chemical species (H{\sc i} or H$_{\alpha}$), $i$ is the cell index, $v_i$ is the velocity of the  
cell, and $\overline{v}$ is the average velocity in the respective direction $k$. $m_j$ is the mass of species $j$ in cell $i$, and $M_{j,\rm{tot}}$ is the total mass of species $j$. We then average over all three directions and receive a mean, one-dimensional velocity dispersion:
\begin{equation}
\sigma_{j, {\rm turb}} = \frac{1}{3}\sum_{k=1}^3 \sigma_{j,k, {\rm turb}}.
\end{equation}

{\sc H{\sc i} velocity dispersion:} For H{\sc i} the intensity is proportional to the number of emitters \citep[e.g.][]{Rohlfs-Wilson96}. Therefore the H{\sc i} mass is a good proxy for the total radiation flux, i.e. $M_{j}\propto L_{j}$, and the resulting velocity dispersion, $\s_{\HI, {\rm turb}}$, can be considered as an intensity-weighted velocity dispersion.

{\sc H$_\alpha$ velocity dispersion:} For H$_\alpha$, the mass of ionised hydrogen is not a good estimate of the H$_\alpha$ intensity, since the emission decreases with increasing temperature. Here, we compute the H$_\alpha$ flux from two contributions, namely the recombination of ionised hydrogen and collisional excitation from the ground state to level $n=3$. We have to neglect the contribution from H{\sc ii} regions around young massive stars since these are not treated in the simulations. Collisional excitations to $n>3$ represent a negligible contribution to the total H$_\alpha$ emission since transitions to these levels are significantly less likely \citep[][but see also \citet{PequignotTsamis05} and references therein]{Anderson+00,Anderson+02}. We compute $m_{\rm{H\alpha}}$ ($dL_{\rm{H\alpha}}$) and $M_{\rm{tot,H\alpha}}$ ($L_{\rm{tot,H\alpha}}$), required in eq. (\ref{sigmaturb}), following the emissivity calculations for recombination and collisional excitation of \cite{Draine11,Dong-Draine11},
and \cite{Kim+13} \citep[but see also][]{Aggarwal83}:
\begin{equation} \label{Lrec}
dL_{\mathrm{H}\alpha,\mathrm{R}}\propto T_4^{-0.942-0.031\ln(T_4)}n_\mathrm{e} n_{\mathrm{H\textsc{ii}}} dV\ ,
\end{equation}
\begin{equation} \label{Lcoll}
dL_{\mathrm{H}\alpha,\mathrm{C}}\propto \dfrac{\Gamma_{13}(T_\mathrm{e})}{\sqrt{T_\mathrm{e}}}e^{\frac{-12.1 \mathrm{eV}}{k_\mathrm{B} T_\mathrm{e}}} 
n_\mathrm{e} n_{\mathrm{H\textsc{i}}} dV\ ,
\end{equation}
where $T_4 = T /10^4$~K, $T_\mathrm{e}$ is the electron temperature, $n_j$ is the number density of species $j$ in
cm$^{-3}$, $dV$ is the zone volume, and
\begin{equation}
  \Gamma_{13}(T_\mathrm{e})=0.35-2.62\times 10^{-7}T_\mathrm{e}-8.15\times
  10^{-11}T_\mathrm{e}^2+6.19\times 10^{-15}T_\mathrm{e}^3.
\end{equation}
We apply equation (\ref{Lcoll}) only for cells with temperatures 4000 $< T < $ 25~000 K assuming $T_\mathrm{e}=T$.
We only consider this temperature range because emission at $T < 4000$ K will be negligible, while we expect to find very little atomic hydrogen at $T > 25~000$~K. We also assume $n_\mathrm{e}=n_{\mathrm{H\textsc{ii}}}$, which results in a $\sim10$\%
error at most in regions where helium is ionised.

We approximate the gas to be optically thin to the H$_\alpha$ and H{\sc i} emission. Although this assumption might lead to a poor estimate of the line intensities, we do not expect major absorption features except in cold and dense H{\sc i} clouds. In this case H{\sc i} self-absorption could also play a non-negligible role. However, as we will see in the next sections, most of the dense clumps in our models are composed of H$_2$, with only a small mass fraction of atomic gas in the outer shells.

{\sc Thermal component:} We include thermal broadening by adding a mass-weighted thermal velocity
\begin{equation} \label{sigmath}
\sigma_{j, \rm{therm}}=\left(\dfrac{\sum_i v_{i,\rm{therm}}^2 \
    m_{i,j}}{M_{j, \rm{tot}}}\right)^{1/2}\ ,
\end{equation}
where
$v_{i, \rm{therm}}=(2 k_B T_i / \mu_i)^{1/2}$ with $T_i$ being the temperature and $\mu_i$ the mean mass per particle in cell $i$. We assume that all species within a cell have the same temperature, which can cause small errors in the thermal velocity estimate. 
The resulting total velocity dispersion is 
\begin{equation} 
\sigma_{{\rm avg\;1D}, j}=\left(\sigma_{j,\rm{turb}}^2+\sigma_{j,\rm{therm}}^2\right)^{1/2}.
\end{equation}

%%%%%%%%%%%%%%%%%%%%%%%%%%%%%%%%%%%%%%%%%%%%

\section{Results and Discussion}\label{sec:results}
%%%%%%%%%%%%%%%%%%%%%%%%%%%%%%%%%%%%%%%%%%%%%%%%%%%%%%%%%%%%%%%%%%%%%%%%%%%%%%%%%%%%%%%%%%%%
\begin{figure}
\includegraphics[width=0.49\textwidth]{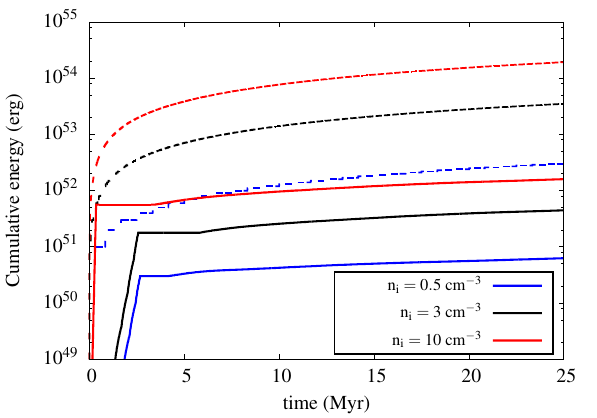}
\caption{Cumulative energy injected in the first 25 Myr of artificial (solid lines, $t=0-25$ Myr) and SN driving (dashed lines, $t=25-50$ Myr) for 
different $n_\mathrm{i}$.}\label{fig:energy}
\end{figure}
\subsection{Energy input}\label{sec:energy input}
In Figure~\ref{fig:energy} we show the cumulative energy input as a function of time for artificial turbulent driving (solid lines) and SN driving (dashed lines) with a SN rate derived from the KS relation, for initial densities $n_\mathrm{i}=0.5, 3$ and $10 \cm$. We point out that the energy input from artificial driving necessary to maintain a constant 3D, mass-weighted, rms
velocity of $v_\mathrm{3D, rms} = 10 \kms$ (corresponding to a one-dimensional velocity dispersion of $\sim 5-6 \kms$) is 
always about two orders of magnitude lower than the energy input in the corresponding SN-driven box. Nevertheless, we will see that SN driving is inefficient in terms of turbulence driving. Therefore, even though the energy input is high, the resulting, average, one-dimensional velocity dispersions in cold H{\sc i} gas stay below $10 \kms$.

%%%%%%%%%%%%%%%%%%%%%%%%%%%%%%%%%%%%%%%%%%%%

\subsection{Impact of supernova positioning}\label{sec:general}
We start from the artificially stirred, turbulent box at different densities at $t_\mathrm{drive}=25$ Myr. We then evolve the simulations with SN driving at different rates and positioning of the SNe relative to the dense gas.

\begin{figure*}
\makebox[1.\textwidth][l]{\large \hspace{2.7cm} Peak driving \hspace{2.7cm} Random driving \hspace{2.7cm} Mixed driving}
\makebox[1.\textwidth][c]{
\begin{overpic}[width=0.3556\textwidth]{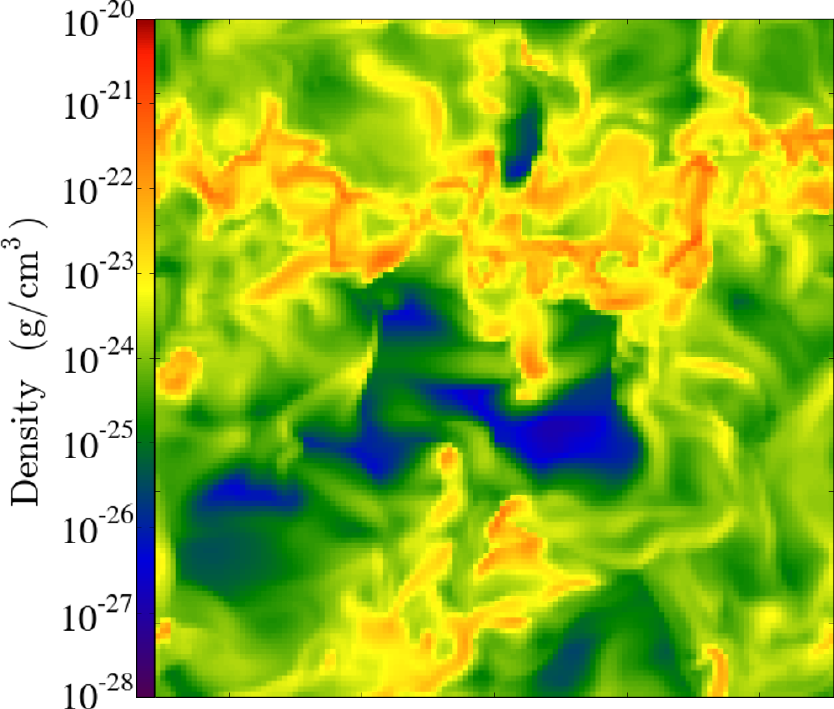}
 \put (58,3) {\textbf{\large \color{black}{time=100 Myr}}}
\end{overpic}
\includegraphics[width=0.29\textwidth]{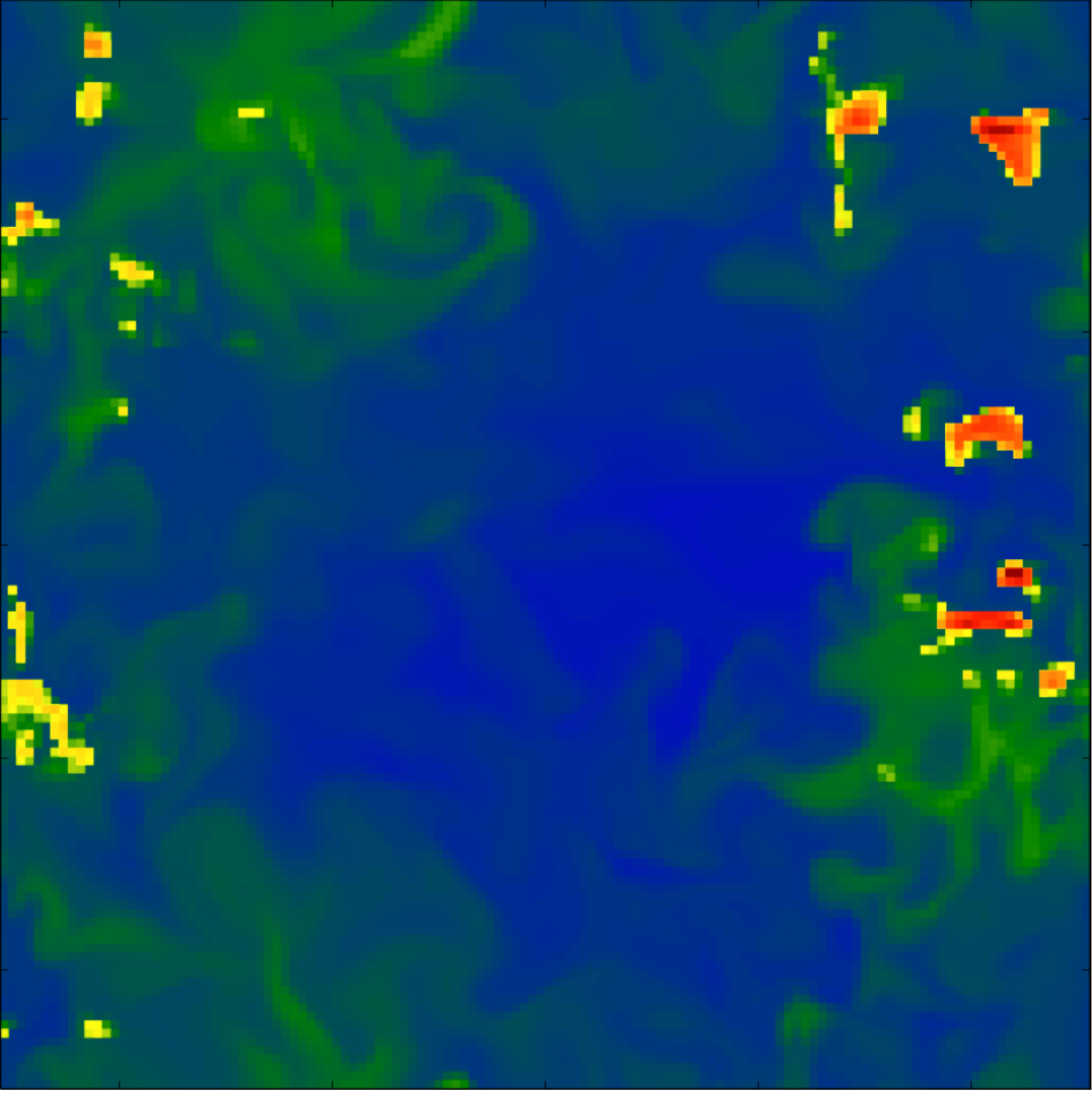}
\includegraphics[width=0.3365\textwidth]{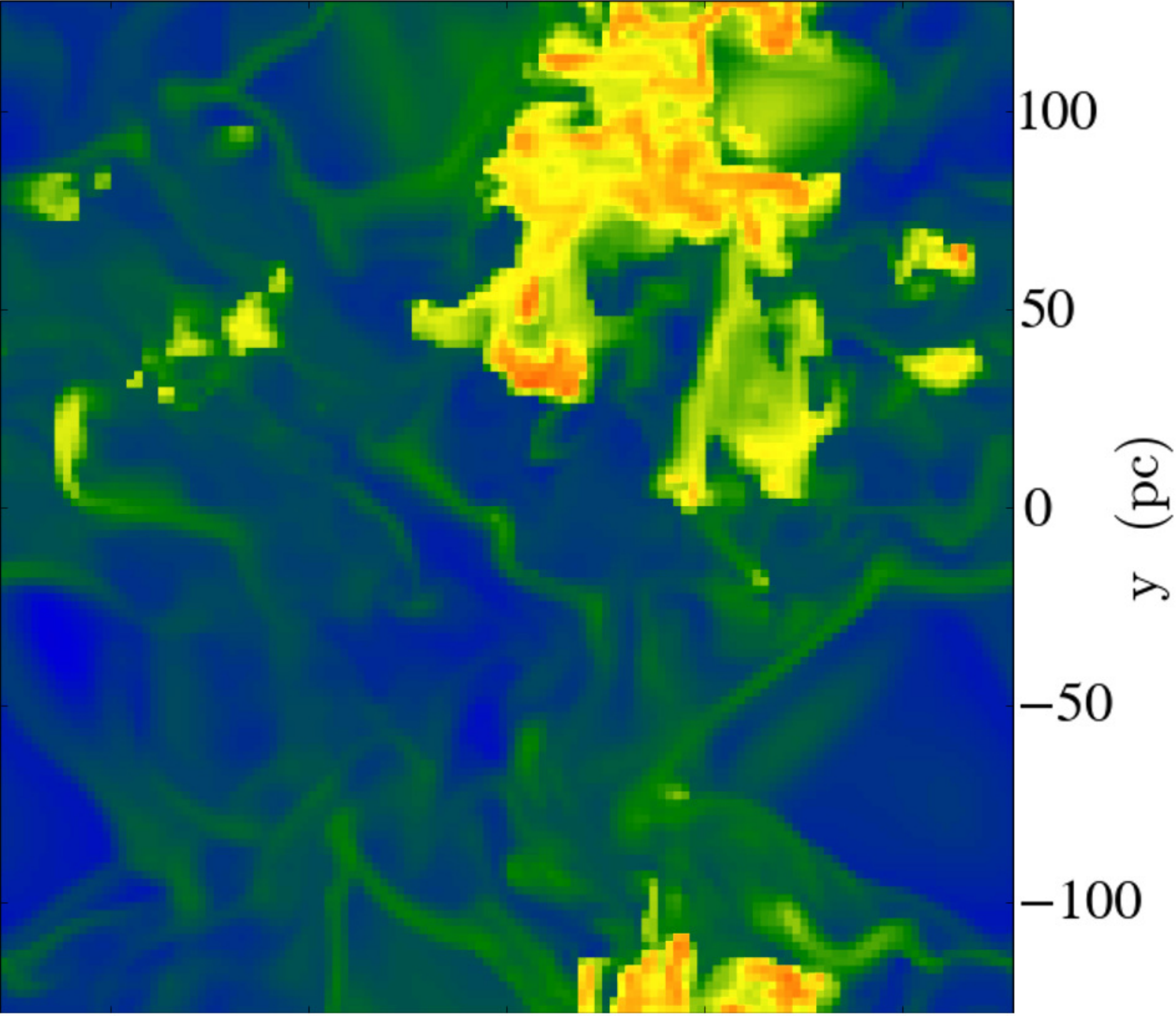}
}
\makebox[1.\textwidth][c]{
\hspace{0.18cm}
\includegraphics[width=0.34\textwidth]{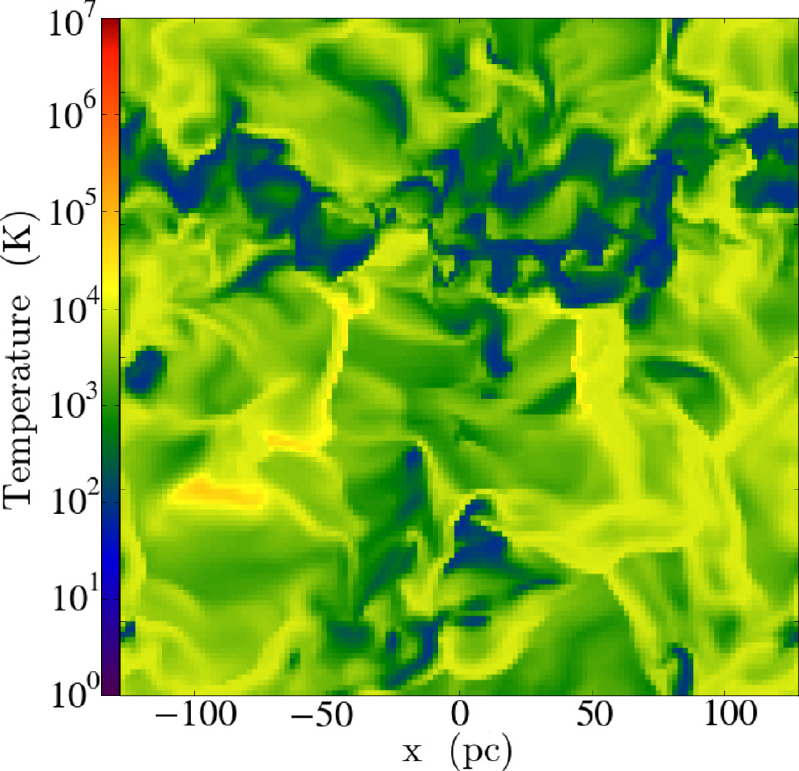}
\includegraphics[width=0.29\textwidth]{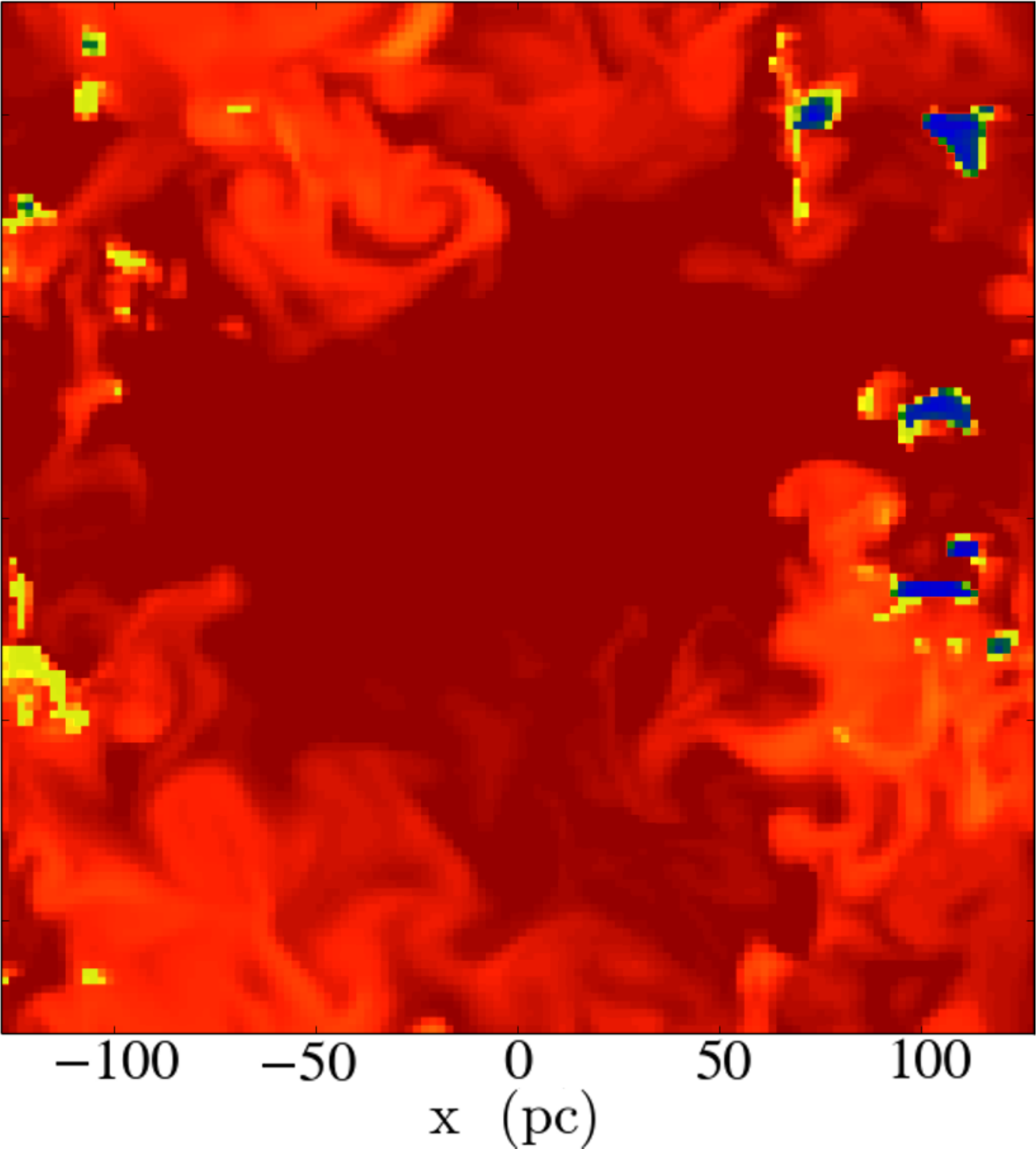}
\includegraphics[width=0.3365\textwidth]{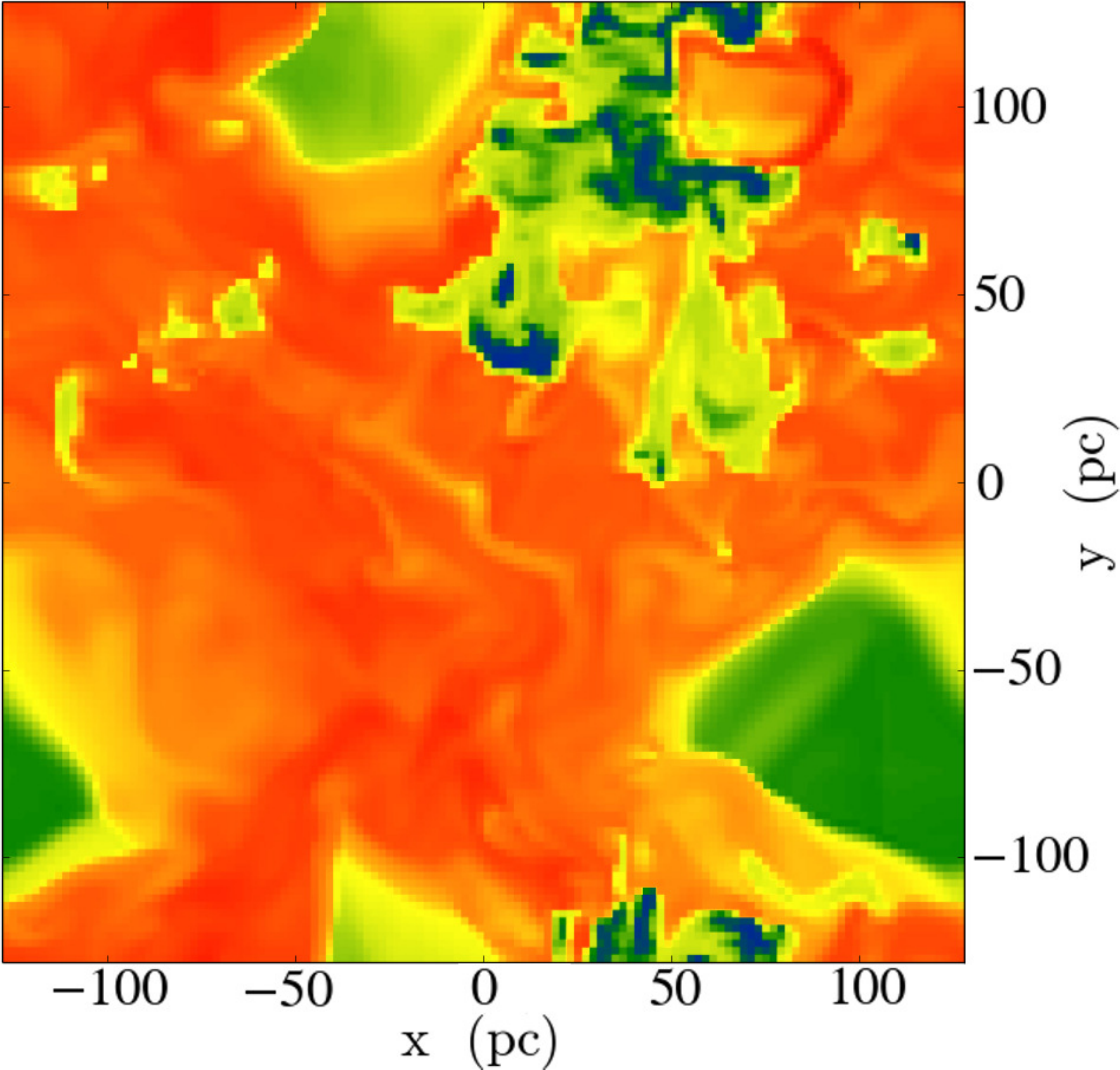}
}
\caption{Density (top row) and temperature (bottom row) slices
for three runs with initial density $n_\mathrm{i}=3 \cm$ and SN rate drawn from the Kennicutt-Schmidt relation $\dot{N}_{\mathrm{SN,KS}}$.
From left to right, we show runs P-n$_\mathrm{3}$ (peak driving), R-n$_\mathrm{3}$ (random driving) and M50-n$_\mathrm{3}$ (mixed driving with $f_\mathrm{peak}=50$\%) at $t=100$ Myr.}\label{fig:slices}
\end{figure*}

{\sc Density and temperature distribution:} ~In Figure~\ref{fig:slices}, we compare the density and temperature structure of the ISM for three runs with $n_\mathrm{i}=3 \cm$ and $\dot{N}_{\mathrm{SN, KS}}$ at $t=100$ Myr, using peak driving (run P-n$_\mathrm{3}$, left column), random driving (run R-n$_\mathrm{3}$, middle column), and mixed driving (run M50-n$_\mathrm{3}$, i.e. mixed driving with $f_\mathrm{peak}=50$\%, right column). The SN rate is the same and the huge differences in the ISM structure are solely due to the positioning of the SNe. For peak driving, the box is filled with warm and cold gas, which is distributed in filaments and extended clouds. There is little to no hot gas present. In the case of random driving, on the other hand, most of the box is filled with hot gas, while the cold 
gas is concentrated in small and dense clumps.
The mixed driving case with an equal contribution from random and peak driving lies in between the other two cases. Gas at $T\gtrsim10^6$ K occupies the majority of the box volume, but the cold and warm phases are located not just in massive clumps, but also in extended filamentary complexes.
%%%%%%%%%%%%%%%%%%%%%%%%%%%%%%%%%%%%%%%%%%%%%%%%%%%%%%%%%%%%%%%%%%%%%%%%%%%%%%%%%%%%%%%%%%%%%
%%%%%%%%%%%%%%%%%%%%%%%%%%%%%%%%%%%%%%%%%%%%%%%%%%%%%%%%%%%%%%%%%%%%%%%%%%%%%%%%%%%%%%%%%%%%%%%%%%%%%%%%%%%%%%%%%%%%%%%%%

\begin{figure}
\includegraphics[width=0.49\textwidth]{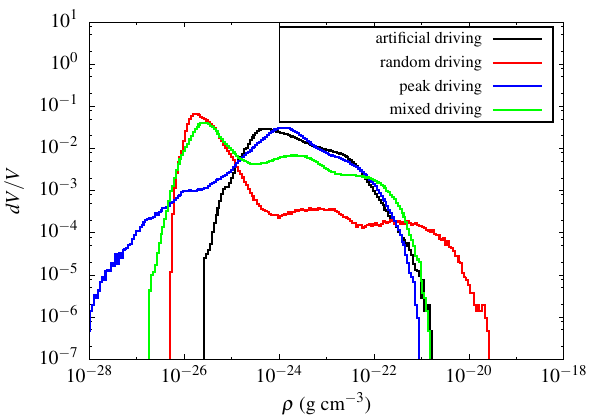}
\caption{Volume-weighted density PDFs for different driving modes, i.e. random (red line), peak (blue line), and 50\% mixed driving (green line), for the n$_{i}=3 \cm$ runs at $t=100$ Myr. In addition, we show the initial condition from artificial stirring at $t_\mathrm{drive}=25$ Myr (black line).}\label{fig:pdf} 
\end{figure}

{\sc Density probability distribution functions:} In Figure~\ref{fig:pdf}, we show the volume-weighted probability distribution functions (PDFs) of density for the initial condition derived from artificial turbulence stirring at $t_\mathrm{drive}=25$ Myr (black line) and for the three different driving modes at $t=100$ Myr.
For artificial driving at a rms velocity of 10 $\kms$, we find a broad PDF which can be interpreted as two overlapping 
log-normal distributions. The medium does not consist of two distinct warm and cold phases.
A significant fraction of the gas lies far from the equilibrium curve due to turbulent motions \citep[e.g.][but see also \citealt{VS+00,Micic+13,Saury+14}]{Walch+11}.
For peak driving (blue line) the PDF roughly coincides with the one for artificial driving at high densities, but with a more extended tail towards low densities. 
In the case of random driving (red line), the distribution has a strong peak at low densities ($\rho \sim {\rm few} \times 10^{-26}\;{\rm g\; cm}^{-3}$), which corresponds to the hot, over-pressured phase, followed by two subsequent peaks at higher densities. The maximum density reached is well above the ones produced
by other driving modes.
Since most of the volume is filled with hot gas, the volume-weighted PDF is dominated by a single peak at low densities.
Mixed driving (green line) displays an intermediate distribution between random and peak driving with three broad peaks. This is typical for a turbulent three-phase ISM consisting of cold, warm and hot gas, with significant contributions from each component \citep[e.g.][]{Gent+13a}.

\begin{figure*}
\includegraphics[width=0.49\textwidth]{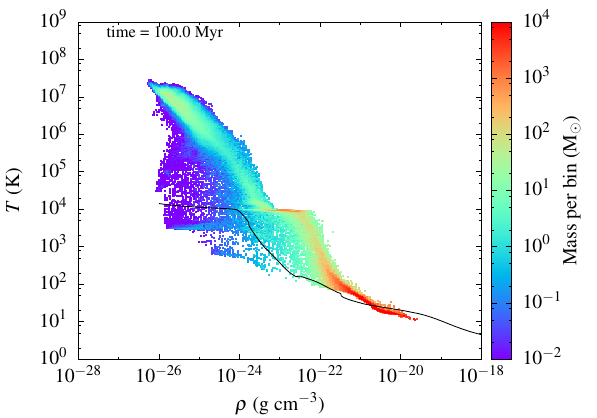}
\includegraphics[width=0.49\textwidth]{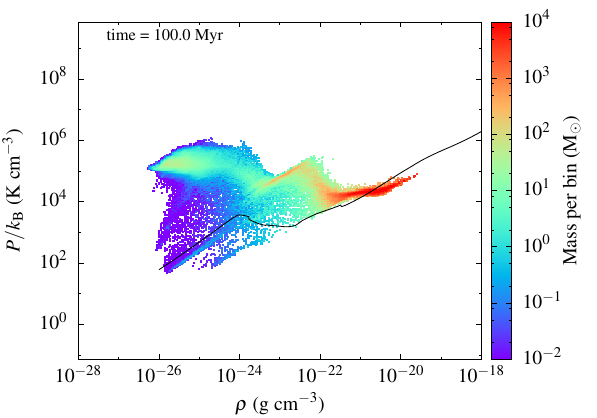}
\includegraphics[width=0.49\textwidth]{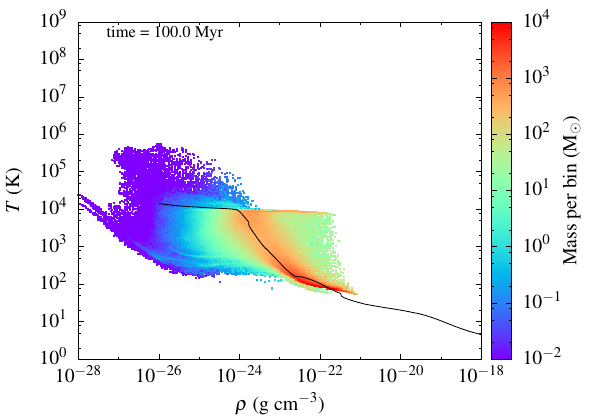}
\includegraphics[width=0.49\textwidth]{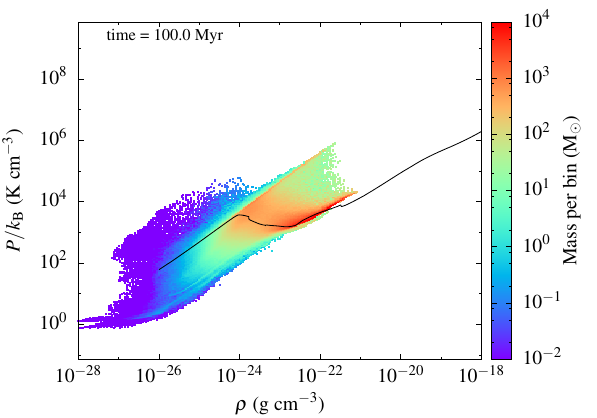}
\caption{Density-temperature (left) and density-pressure (right) phase diagrams for two simulations with initial density $n_\mathrm{i}=3 \cm$ and $\dot{N}_{\mathrm{SN, KS}}$ at $t=100$ Myr. We colour-code the mass distribution. The first row shows run R-n$_\mathrm{3}$ (random driving), while the second row shows run P-n$_\mathrm{3}$ (peak driving). The solid lines show where radiative cooling and diffuse radiative heating reach equilibrium.}\label{fig:phase-diagrams}
\end{figure*}

{\sc Phase diagrams:} In Figure~\ref{fig:phase-diagrams}, we show exemplary $\rho-T$ and $\rho-P$ phase diagrams for random and peak driving, where the mass distribution is colour-coded. In the random driving case (top row) the gas is moved to high pressures $P/k_B \gtrsim 10^4\;{\rm K\; cm}^{-3}$, which is above the mean pressure in the Milky Way \citep[e.g.][]{JenkinsTripp11}. A small fraction of the gas reaches high temperatures $T > 10^6$~K; this gas occupies virtually all of the volume, as shown in Figure~\ref{fig:pdf}.
This hot, high-pressure gas pushes the majority of the mass into the cold, dense phase (see section \ref{sec:thermal}).
The cold, high density tail lies slightly below the equilibrium curve. A likely explanation is that at high density, the cold gas is efficiently shielded against the far-UV interstellar radiation field. Since the equilibrium values are computed without considering this mechanism, the densest component has a lower equilibrium pressure and temperature in this density range.

On the other hand in the peak driving case (bottom row) a two-phase medium, with warm and cold gas in approximate pressure equilibrium, is formed. The densest component actually lies slightly above the equilibrium curve due to the continuous heating of the densest gas parcels by SN shocks. However, these simulations contain little to no hot gas, as the SNe, which explode only in dense environments, fail to effectively heat the gas (see section \ref{sec:peak}).  The maximum densities reached in simulations with peak driving are 1--2 orders of magnitude smaller than in the random driving case since, by construction, peak driving removes the densest component of the medium and highly over-pressured hot regions are not present in this case.
%%%%%%%%%%%%%%%%%%%%%%%%%%%%%%%%%%%%%%%%%%%%
%%%%%%%%%%%%%%%%%%%%%%%%%%%%%%%%%%%%%%%%%%%%
%%%%%%%%%%%%%%%%%%%%%%%%%%%%%%%%%%%%%%%%%%%%
%%%%%%%%%%%%%%%%%%%%%%%%%%%%%%%%%%%%%%%%%%%%

{\sc Summary:} We may summarise the impact of the different SN positioning (see Figs. \ref{fig:slices} and \ref{fig:phase-diagrams}) as follows: 
in the case of random driving, most SNe are placed in positions where their energy can be efficiently transferred to the ISM. In the case of peak driving, the SNe explode in dense gas, which may promptly radiate a large fraction of the inserted energy. The higher the fraction of random SNe, the more efficiently energy is injected, which increases the thermal gas pressure and increases the chance to end up in a thermal runaway regime, as discussed in the next subsection. In section \ref{sec:mixed} we explore this transition in simulations with different $f_\mathrm{peak}$.

\subsection{ Thermal runaway}\label{sec:thermal}
The high pressure reached in our models results in the equilibrium between heating and cooling lying not in the regime near $P/k_B \sim 10^3\;{\rm K\; cm}^{-3}$ where a two-phase medium is possible, but rather on the cold branch of the equilibrium curve, where balance between heating and cooling can only be reached at low temperatures and high densities.
This explains why virtually all the mass lies at high density ($n > 100$~cm$^{-3}$) and low temperature ($T < 200$~K).
As a result, we move from the classic paradigm of a two-phase medium in pressure equilibrium towards a scenario in which only the cold phase survives, consistent with the picture of \citet{Wolfire+95} for high heating (far-UV interstellar field) rates. The cooling times in the remaining rarefied gas are long, since radiative cooling is proportional to $n^2$ and drops significantly for temperatures above $10^6$~K \citep{Raymond+76}.
As a result, almost all of the volume is occupied by gas with low density and $T > 10^6$~K, whose cooling time far exceeds the dynamical time of the system: the third phase of the three-phase medium \citep{McKeeOstriker77}.
This picture is consistent with the findings of \cite{Scannapieco+12}, who use hydrodynamical simulations of artificially-driven turbulence in a stratified disc to mimic the impact of stellar feedback. They find that, for turbulent one-dimensional velocity dispersions $\gtrsim 35 \kms$, large fractions of gas are continuously heated and unable to cool within a turbulent crossing time. This process leads to powerful outflows and compression of cold gas. In the thermal runaway regime, our simulations behave in a similar way, as energy is directly deposited into a hot, high pressure medium, with a resulting one dimensional velocity dispersion of the hot phase well above the critical value of $35 \kms$.

The high pressures reached in the thermal runaway regime marks a clear distinction from the classic picture of \citet{McKeeOstriker77}. In that model, a SN remnant expands into a medium characterised by the presence of many evenly-distributed, small, two-phase clouds having cold cores surrounded by warm envelopes. The passage of the SN blast wave heats and destroys these clouds, sweeping them up into a dense shell. This dense shell cools and slows down during the radiative phase, and the remnant ceases to exist when it reaches pressure equilibrium with the ambient medium.
For thermal runaway, however, SNe go off in a hot, high pressure environment, while isolated cold clouds lie far from the explosion position.
Well before the end of the Sedov-Taylor phase, the remnants reach pressure equilibrium with the hot gas and deposit their energy into the medium without any significant radiative loss.
\begin{figure}
\includegraphics[width=0.49\textwidth]{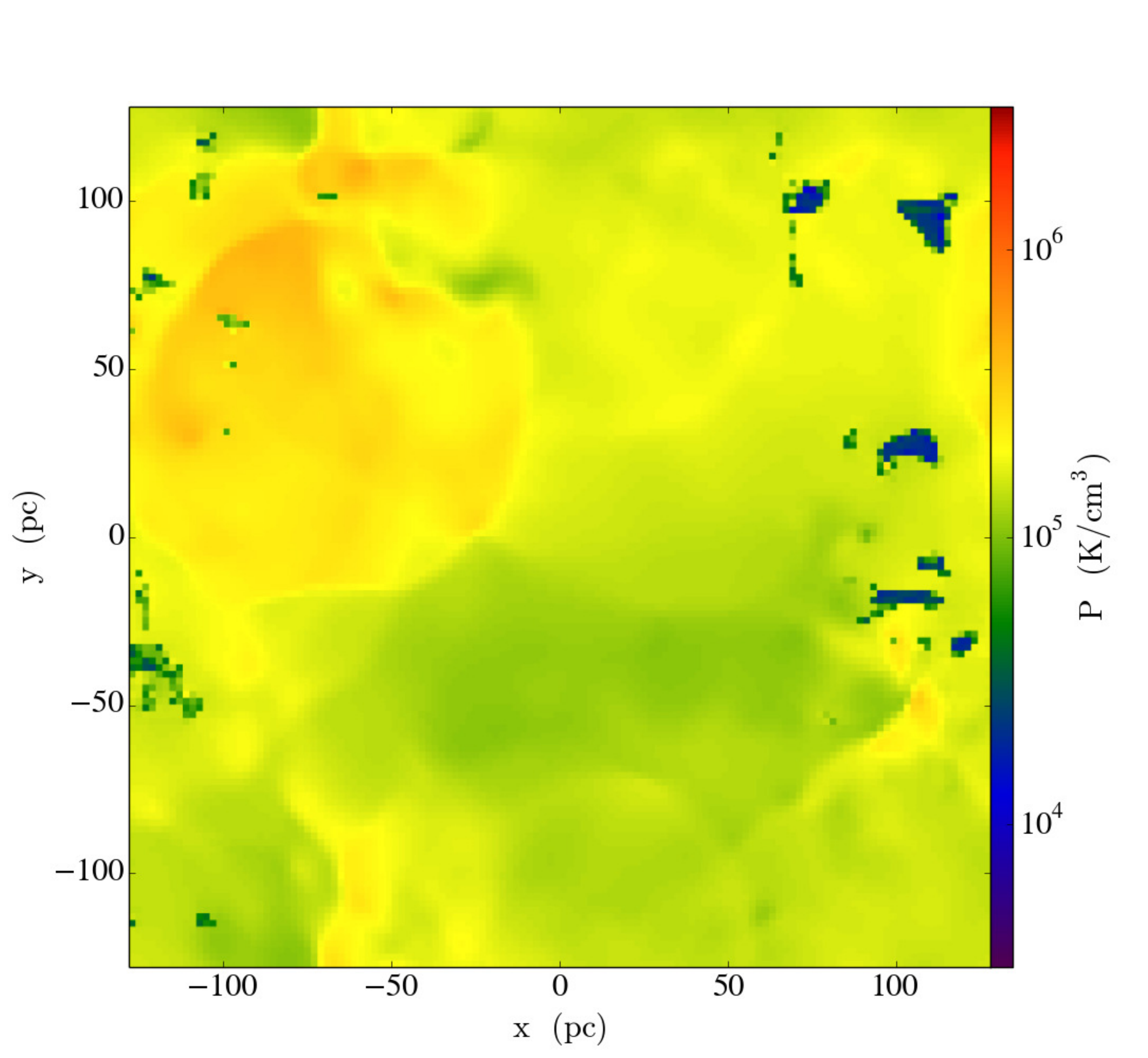}
\caption{Pressure slice for the random driving run with initial density $n_\mathrm{i}=3 \cm$ and $\dot{N}_{\mathrm{SN,KS}}$ at $t=100$ Myr (see also central column of Fig.\ \ref{fig:slices}).}\label{fig:pslice} 
\end{figure}
%%%%%%%%%%%%%%%%%%%%%%%%%%%%%%%%%%%%%%%%%%%%%%%%
%%%%%%%%%%%%%%%%%%%%%%%%%%%%%%%%%%%%%%%%%%%%
\begin{figure*}
\includegraphics[width=0.49\textwidth]{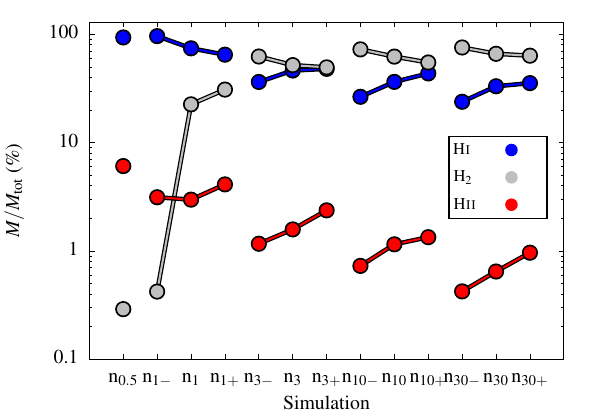}
\includegraphics[width=0.49\textwidth]{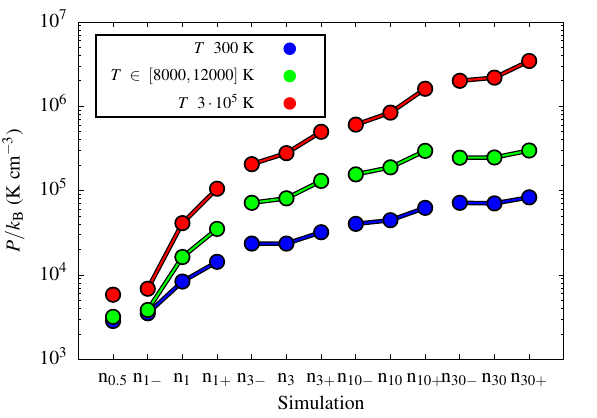}
\includegraphics[width=0.49\textwidth]{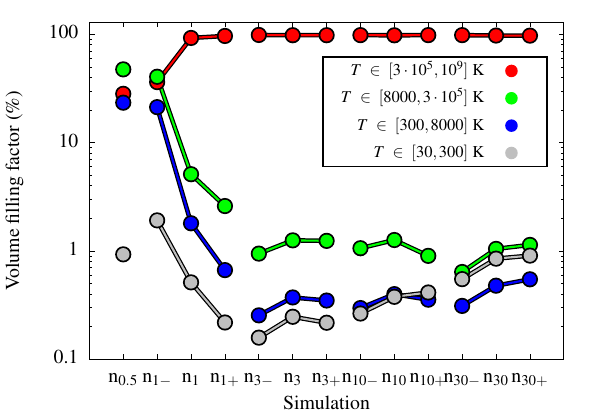}
\includegraphics[width=0.49\textwidth]{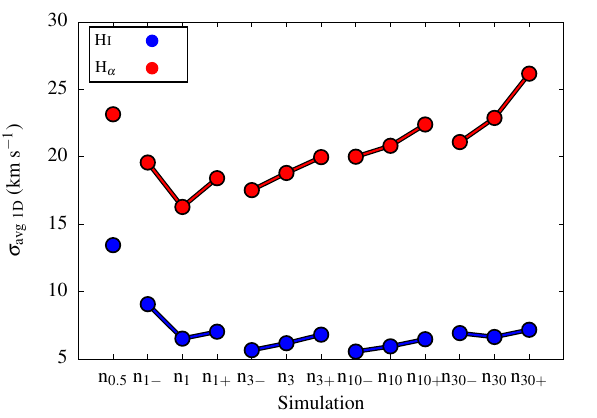}
\caption{Average mass fractions (top left), pressures (top right), VFFs (bottom left) and 1D velocity dispersions (bottom right) with random driving for different initial densities $n_\mathrm{i}$ and SN rates. The index $i$ represents the initial number density,
while the $+$ and $-$ signs give the SN rate ($\dot{N}_\mathrm{SN,+}$ and $\dot{N}_\mathrm{SN,-}$).
The values are averaged over the last 5 Myr (section \ref{sec:sims}). The different lines connect simulations with same $n_\mathrm{i}$ but different SN rate.}\label{fig:compare-random}
\end{figure*}
%%%%%%%%%%%%%%%%%%%%%%%%%%%%%%%%%%%%%%%%%%%%

In this environment heating cannot be balanced by radiative cooling. In our periodic box, within which there is no vertical stratification to allow an atmosphere to evolve into hydrostatic equilibrium with a galactic potential, the hot gas pressure can only be set by the balance between cooling and heating. Random driving thus pushes the system towards a thermal runaway regime, where the hot gas contains most of the energy, while the majority of the box mass is stored in the coldest phase without the possibility to return to a system with a significant warm phase ($T\sim10^4$ K).
Cold gas is not formed from a two-phase medium, but instead cools directly to the cold branch of the equilibrium curve, forming high-density clumps with small VFF. The cold clumps are produced by compression waves coming from the hot, high-pressure gas. This mechanism causes an efficient and fast conversion from warm, atomic gas to molecular.
These clumps are resolved by so few zones that they cannot reach high enough densities to remain in pressure equilibrium with the background. As a result, cold clouds at low pressures are embedded into a hot, high-pressure environment. This numerically caused jump in pressure can reach about one order of magnitude (see Fig.\ \ref{fig:pslice} for the pressure 
distribution in run with $n_\mathrm{i}=3 \cm$ and $\dot{N}_{\mathrm{SN,KS}}$). Therefore, we must treat the interior properties of the clouds as limits at low density and velocity dispersion, rather than converged values.

\subsection{Random driving at different SN rates}\label{sec:random}

In this section we discuss the properties of the ISM forming in runs R-n$_\mathrm{i}$ with randomly placed SNe and different initial densities $n_\mathrm{i}$. The SN rates are adjusted to the given $n_\mathrm{i}$ (see section \ref{sec:SNdriving}). We perform three simulations with different SN rates for almost every $n_\mathrm{i}$ using $\dot{N}_\mathrm{SN,KS}$, and $\dot{N}_\mathrm{SN,-}$ and $\dot{N}_\mathrm{SN,+}$, where the SN rate is decreased/increased by a factor of two, respectively (see also Table \ref{tab:nSNR}). In Figure~\ref{fig:compare-random} we compare the different simulations. We plot the mass in ionised, atomic, and molecular hydrogen (top left panel), the pressure in three different temperature regimes, which are representative for the cold, warm ionised, and hot phase of the ISM (top right), the VFF in different temperature phases (bottom left), and the velocity dispersion of the gas in H{\sc i} and H$_\alpha$ (bottom right) towards the end of each simulation.
To guide the eye we connect the simulations with equal $n_\mathrm{i}$ but different SN rate.\\

{\sc Chemical composition:}
We follow the chemical evolution of the gas, including the formation of molecular hydrogen, taking into account the effects of dust shielding and molecular (self-)shielding. We find that the mass in ionised hydrogen is always below 10\% and decreasing for increasing box density. For low densities ($n_\mathrm{i}\lesssim3 \cm$) most of the total mass is in atomic hydrogen. At higher densities ($n_\mathrm{i}\gtrsim3 \cm$) less than 50\% of the mass is in H{\sc i} and the rest is in molecular hydrogen. Molecular hydrogen, which is organised in small, dense clumps (see Fig.\ \ref{fig:slices}), dominates the mass budget at high densities and SN rates (up to 80\% of the total mass is in form of H$_2$ at $n_\mathrm{i}=30 \cm$).  

{\sc Gas pressure:} We compute the average pressure in three different temperature regimes: (i) for the stable cold phase at $T \leqslant 300$ K; (ii) for the stable warm phase at $8000\leqslant T \leqslant12000$ K; and (iii) for the hot phase at $T \geqslant3\cdot 10^5$ K. For $n_\mathrm{i}\leqslant1 \cm$ the medium is roughly in pressure equilibrium, but for higher densities and SN rates the pressures of the three phases diverge slowly as we get into the thermal runaway regime.

{\sc Volume Filling Fractions:}
We show the VFFs of the gas in the bottom left panel of Figure~\ref{fig:compare-random}. We distinguish four different temperature regimes: the cold phase at $30\leqslant T < 300$ K; the warm atomic ISM at $300\leqslant T <8000$ K; the warm ionised medium at $8000\leqslant T <3\times 10^5$~K; and the hot ionised medium at $T \geqslant3\times 10^5$~K. Only for two simulations (R-n$_\mathrm{0.5}$ and R-n$_\mathrm{1-}$), are the VFFs close to what we expect for a Milky Way-type galaxy near the mid-plane \citep[e.g.][]{KalberlaDedes08,KalberlaKerp09}, whereas the hot gas fills most of the volume at higher densities and/or SN rates.

We compare the VFFs of the hot ionised medium with the analytic prediction of \cite{McKeeOstriker77}. The volume occupied by SN remnants at random locations in a uniform medium, $f$, can be written as
\begin{equation} \label{fSNR}
f = 1-e^{-Q}\ ,
\end{equation}
with porosity $Q$ defined as
\begin{equation} \label{QSNR}
Q = 10^{-0.29} E_{\mathrm{51}}^{1.28} S_{\mathrm{-13}} \bar{n}^{-0.14} \tilde{P}_{\mathrm{04}}^{-1.3}\ ,
\end{equation}
where $E_{\mathrm{51}}$ is the SN energy normalised to $10^{51}$ erg, $S_{\mathrm{-13}}$ is the SN rate in units of $10^{-13}$ pc$^{-3}$ yr$^{-1}$, $\bar{n}$ is the number density of the ambient medium in $\cm$, and $\tilde{P}_{\mathrm{04}} = 10^{-4}P_{\mathrm{0}}/k_{\mathrm B}$, with $P_{\mathrm{0}}$ and $k_{\mathrm B}$ ambient medium pressure and Boltzmann constant, respectively.

We find a reasonable agreement of the VFF for models R-n$_\mathrm{0.5}$ and R-n$_\mathrm{1-}$, where $f\sim$ 30\% - 40\%. However, the analytic model predicts lower hot gas VFFs for higher densities and SN rates such than the ones found in our higher SN rate simulations. Since $Q\propto \bar{n}^{1.26}  \tilde{P}_{\mathrm{04}}^{-1.3}$, $f$ as derived from eq. (\ref{fSNR}) and (\ref{QSNR}) is roughly constant, whereas the simulation shows it increasing towards unity in the thermal runaway regime.

{\sc H{\sc i} and H$_\alpha$ velocity dispersion:}
In the bottom right panel of Figure~\ref{fig:compare-random}, we show the velocity dispersions in H{\sc i}, $\sigma_\mathrm{avg\;1D}$(H{\sc i}), and H$_\alpha$, $\sigma_\mathrm{avg\;1D}($H$_{\alpha})$. For $n_\mathrm{i}\le 1 \cm$ and, in particular, for low SN rates ($\dot{N}_\mathrm{SN}\lesssim 2$ Myr$^{-1}$), SNe are able to inject the observed level of turbulence in the H{\sc i} gas, while $\sigma_\mathrm{avg\;1D}(\HI) \sim 5-7 \kms$ for $n_\mathrm{i}>1 \cm$. Due to the small size of most cold clumps, these values are a mixture of the clump-to-clump velocity dispersion and thermal broadening, rather than being a measure of the disordered motions within these clouds. 
For the H$_\alpha$ dispersion, we find a similar trend going from SN rates of $\dot{N}_\mathrm{SN}\lesssim 2$ Myr$^{-1}$ to slightly higher ones. First $\sigma_\mathrm{avg\;1D}($H$_{\alpha})$ drops from $\sim 24$ $\kms$ to $\sim 17$ $\kms$, but then it increases slowly with increasing box density and SN rate and is $\sim 26$ $\kms$ for $n_\mathrm{i}=30\;\cm$ and high SN rates.

\subsection{Peak driving vs. random driving}\label{sec:peak}
%%%%%%%%%%%%%%%%%%%%%%%%%%%%%%%%%%%%%%%%%%%%
\begin{figure*}
\includegraphics[width=0.49\textwidth]{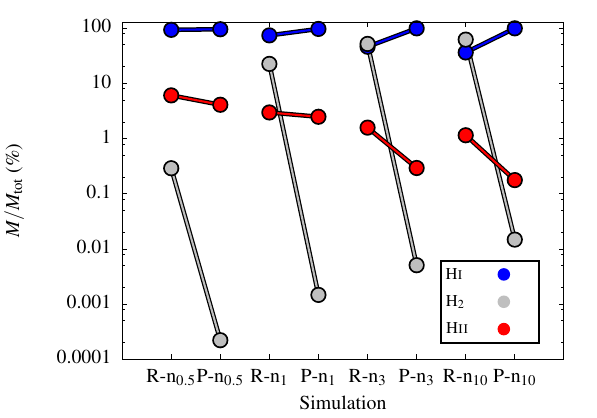}
\includegraphics[width=0.49\textwidth]{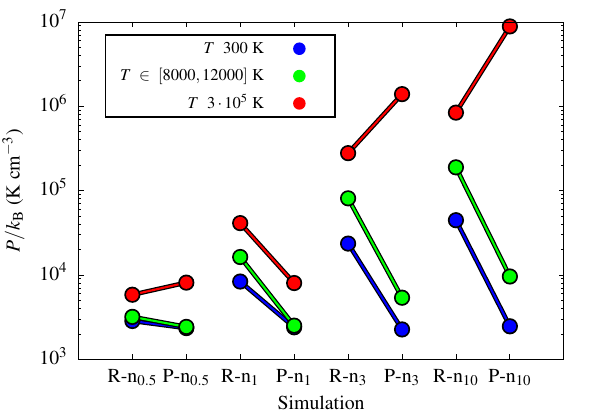}
\includegraphics[width=0.49\textwidth]{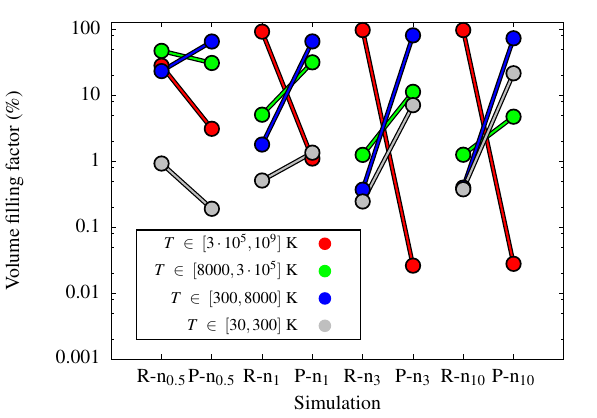}
\includegraphics[width=0.49\textwidth]{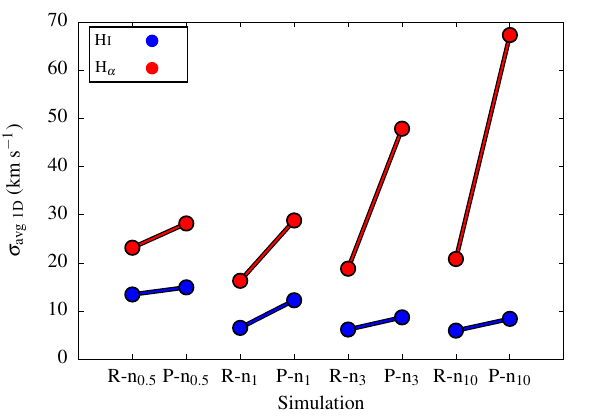}
\caption{Average mass fractions (top left), pressures (top right), VFFs (bottom left) and 1D velocity dispersions (bottom right) with random (R) and peak (P) driving for different initial densities $n_\mathrm{i}$, using the corresponding $\dot{N}_{_{\rm SN, KS}}$. The lines connect simulations with equal $n_\mathrm{i}$ and SN rate but different driving modes (R or P).}\label{fig:compare-peak}
\end{figure*}
%%%%%%%%%%%%%%%%%%%%%%%%%%%%%%%%%%%%%%%%%%%%
We perform simulations using peak driving (runs P-$n_\mathrm{i}$) at the same box densities and SN rates as in the corresponding runs with random driving (see Table \ref{tab:runs}). In Figure~\ref{fig:compare-peak} we show the resulting ISM properties with respect to the random driving case.

{\sc Chemical composition:} The peak driving runs are dominated by atomic rather than molecular gas. The H$_2$ mass fraction is low for two reasons: (i) the densest peaks are dispersed by SNe and, therefore, H$_2$ is locally dissociated; and (ii) the strong compressive SN shocks, which are widespread in the case of random driving, are localised in the immediate neighbourhood of the SN for peak driving.
The destruction of molecular gas from peak SNe is broadly consistent with the findings of \cite{HennebelleIffrig14} and \cite{IffrigHennebelle14}, where SNe exploding close to or within star formation sites can disperse cold and dense gas.
In addition, the inefficient heating by peak SNe further reduces the small fraction of mass in the form of thermally ionised hydrogen.

{\sc Gas pressure:} We find lower pressures for peak driving, in particular for the warm and cold gas components.  However, the pressure of the hot phase ($T \ge 3\times 10^5$ K) appears to be significantly higher (by a factor of $\sim$ 10) in case of peak driving at $n_\mathrm{i} \ge 3\;{\rm cm}^{-3}$. This increase has to be interpreted in conjunction with the decreasing VFF of the hot gas at these densities. Since the SNe, which explode in high density environments, are subject to strong radiative cooling, the high pressure reflects the young age of the SN remnants that do contribute to the hot phase.

{\sc Volume Filling Fractions:} Peak driving does not produce a predominantly hot ISM. Instead, the VFF is highest for the warm and cold atomic phases. In particular for low densities, a non-negligible contribution is also produced by the warm ionised medium, but the VFF of this component decreases with increasing density, similar to the case of random driving. In general, the absence of a significant hot phase reflects the small sphere of influence of each peak SN, whose expansion is stopped early on due to strong radiative cooling.

{\sc H{\sc i} and H$_\alpha$ velocity dispersion:} In the case of peak driving, the $\HI$ velocity dispersion is slightly higher than for random driving. This is reasonable since the SNe deposit momentum into the cold gas. The velocity dispersion of the warm ionised gas as seen in H$_{\alpha}$ grows even more significantly up to $50-70$ $\kms$ for $n_\mathrm{i} \ge 3\;{\rm cm}^{-3}$. As we have seen, the hot and warm ionised gas are found in the earliest stages of the SN remnants, which otherwise cool efficiently. Thus, the apparent high values of the H$_{\alpha}$ velocity dispersion stem from the integration over a number of isolated, compact, young SN remnants.

\subsubsection{Discussion: Is peak driving realistic?}\label{sec:overcool}
Figure~\ref{fig:compare-peak} shows that peak driving efficiently disperses cold gas. The SNe which, by choice, explode in the densest environments, are subject to strong radiative cooling. 
For this reason, peak driving fails to reproduce both the VFF of hot gas and the large molecular gas mass fractions characteristic of the Milky Way \citep{Ferriere01}. The absence of hot gas is also inconsistent with the expectation of SNe being responsible for the creation of a hot phase \citep{McKeeOstriker77}.
  
We can conclude that pure peak driving does not reproduce realistic ISM conditions -- probably because we neglect other important physical ingredients, such as clustering and stellar feedback mechanisms, i.e. pre-SN feedback like stellar winds and ionising radiation \citep{Walch+12-2}. 
%In particular, the early feedback of massive stars modifies the environment of the SN progenitor \citep{WalchNaab2014} and it might be important to include these processes when modelling a SN-driven ISM, where the SNe are placed in the densest gas.
However, one has to be cautious not to over-estimate the effect of radiative cooling due to the finite numerical resolution of our models. Numerical over-cooling acting at the interface between the cold shell and the hot interior of a SN bubble may reduce the amount of hot gas. More importantly, the density and, therefore, the mass within the SN injection region is high in case of peak driving. Since we always inject a SN energy of $10^{51}$ erg per explosion, the effective temperature within the injection region can drop below 10$^6$ K in a dense environment. This is an unfavourable temperature regime, the cooling curve is steep and the heated SN gas can be cooled efficiently.

\begin{figure}
\includegraphics[width=0.49\textwidth]{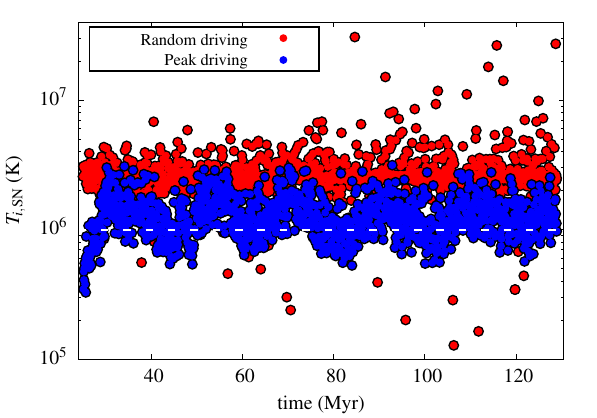}
\caption{Initial temperature of SNe estimated from equation (\ref{TiSN}) for pure peak and random driving with $n_\mathrm{i}=3 \cm$ and 
$\dot{N}_{_{\rm SN, KS}}$.}\label{fig:TiSN}
\end{figure}

Figure~\ref{fig:TiSN} shows the approximate initial temperatures (from eq. \ref{TiSN}) of all SNe in case of $n_\mathrm{i}=3 \cm$ and $\dot{N}_{\mathrm{SN, KS}}$ for run R-$n_3$ with random and run P-$n_3$ with peak driving. The temperature within the injection region, $T_{\mathrm{i, SN}}$, depends on the density. Typically, the temperature is $T_{\mathrm{i, SN}} > 2 \times 10^6$ K in case of random driving. For peak driving we find $T_{\mathrm{i, SN}} \gtrsim 10^6$ K, and for $\sim$35\% of all explosions the temperature is below $10^6$ K, leading to immediate strong cooling.
For this reason, we explore a combined energy and momentum input model in section \ref{sec:momentum}.
%%%%%%%%%%%%%%%%%%%%%%%%%%%%%%%%%%%%%%%%%%%%
%%%%%%%%%%%%%%%%%%%%%%%%%%%%%%%%%%%%%%%%%%%%

%%%%%%%%%%%%%%%%%%%%%%%%%%%%%%%%%%%%%%%%%%%%

\subsection{Mixed driving}\label{sec:mixed}
%%%%%%%%%%%%%%%%%%%%%%%%%%%%%%%%%%%%%%%%%%%%
\begin{figure*}
\includegraphics[width=0.49\textwidth]{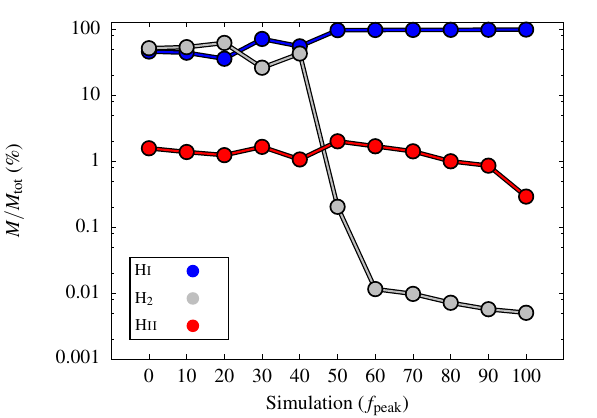}
\includegraphics[width=0.49\textwidth]{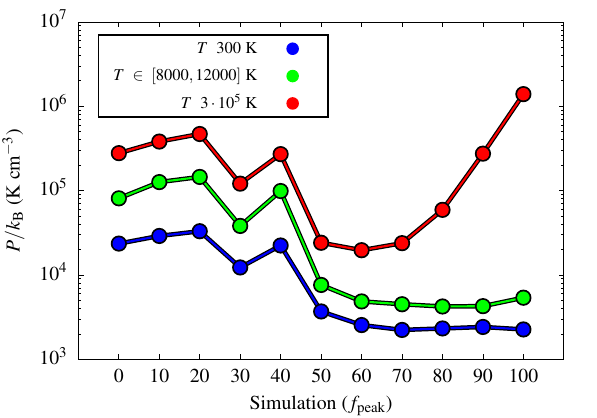}
\includegraphics[width=0.49\textwidth]{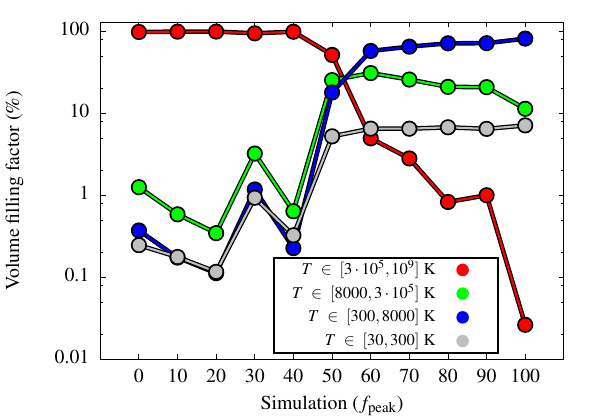}
\includegraphics[width=0.49\textwidth]{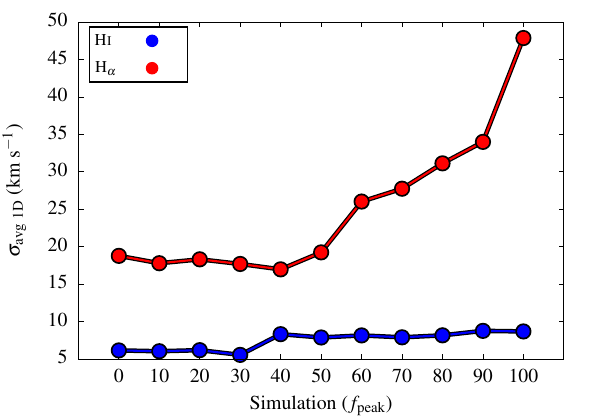}
\caption{Average mass fractions (top left), pressures (top right), VFFs (bottom left) and 1D velocity dispersions (bottom right) as a function of $f_\mathrm{peak}$ for simulations using $n_\mathrm{i}=3 \cm$ and $\dot{N}_{\mathrm{SN, KS}}$.}\label{fig:compare-mixed}
\end{figure*}
%%%%%%%%%%%%%%%%%%%%%%%%%%%%%%%%%%%%%%%%%%%%
Peak and random driving represent two extreme cases of what we expect for the spatial distribution of SNe.
To examine intermediate cases, we perform 9 additional simulations with $n_\mathrm{i} = 3 \cm$ and $\dot{N}_{\mathrm{SN, KS}}$ and different values of $f_\mathrm{peak}$, the fraction of SNe going off in dense gas. Here, $f_\mathrm{peak}=0$\% corresponds to random driving and $f_\mathrm{peak}=100$\% corresponds to peak driving. In Figure~\ref{fig:compare-mixed} we show the mass fractions (top left), pressures (top right), VFFs (bottom left) and 1D velocity dispersions (bottom right) as a function of $f_\mathrm{peak}$.

{\sc Chemical composition:} The ratio of atomic to molecular gas has a steep transition at $f_\mathrm{peak}\approx50$\%. For $f_\mathrm{peak}>50$\% the box is H{\sc i}-dominated, which is typical for an ISM where the coldest component is removed by SN explosions within the dense gas. On the other hand, for $f_\mathrm{peak}<50$\%, we find large amounts of molecular hydrogen since (i) a smaller $f_\mathrm{peak}$ disperses fewer dense clumps, and (ii) the larger number of uncorrelated SNe heat up the gas around the dense and cold medium and compress it (see section \ref{sec:thermal}).

{\sc Gas pressure:} We find that all three phases are out of pressure equilibrium, although the pressures of the cold and warm phases decrease with increasing $f_\mathrm{peak}$ and these phases become close to isobaric for $f_\mathrm{peak}>50$\%. However, the hot gas pressure diverges for $f_\mathrm{peak}>50$\% as the SN remnants which contribute to this phase become younger and occupy smaller volumes. This has already been discussed for the case of pure peak driving
(see section \ref{sec:peak}). 

{\sc Volume Filling Fractions:} A sharp transition at $f_\mathrm{peak}\approx50$\% can also be found for the VFFs of the different gas phases. For $f_\mathrm{peak}\gtrsim50$\%, the hot gas VFF drastically decreases while the VFFs of the warm and cold components are increasing. For small $f_\mathrm{peak}<50$\% the number of random SN explosions is high enough to fill most of the box with hot gas and drive the box towards the thermal runaway regime.

{\sc H{\sc i} and H$_\alpha$ velocity dispersion:}
With increasing peak fraction, the $\HI$ velocity dispersion increases from $\sim 5-6$ $\kms$ to $\sim8-9$ $\kms$ at $f_\mathrm{peak}\approx 40$\%. 
Also the H$_{\alpha}$ velocity dispersion increases for $f_\mathrm{peak}\gtrsim 40$\%, from $\sim 18$ $\kms$ to $\sim 48$ $\kms$ for $f_\mathrm{peak}=100$\%. For high $f_\mathrm{peak}$, most of the H$_\alpha$-emitting gas comes from young SN remnants (see section \ref{sec:peak}).
%This is potentially interesting for high-redshift, gas-rich galaxies, which are observed to have high $\sHa > 50$ $\kms$ \citep{Genzel+11} and high SFRs.

\subsubsection{Discussion: the transition between the peak and random driving regime.}
Mixed driving, i.e. a combination of peak and random driving at different ratios, shows a relatively sharp transition in most ISM properties at a critical ratio of $f_{\rm peak, crit}\approx 50$\%. It is likely that $f_{\rm peak, crit}$ depends on the box density and SN rates. For instance, we expect $f_{\rm peak, crit}$ to shift to higher values for higher average densities. It is also likely that $f_{\rm peak, crit}$ shifts towards lower values if the periodic boundary conditions are relaxed and the box is allowed to 'breathe' (i.e. to adjust to local pressure equilibrium as gas is allowed to escape the box). In this case the thermal runaway can also be delayed, leading to small $f_{\rm peak, crit}$, or even avoided altogether. 
Thus, rather than being interested in extrapolating detailed physical conclusions, we are more keen on stressing once again that there are major differences in the properties of the ISM, which results from implementations of either peak or random driving.
%%%%%%%%%%%%%%%%%%%%%%%%%%%%%%%%%%%%%%%%%%%%
\begin{figure*}
\includegraphics[width=0.49\textwidth]{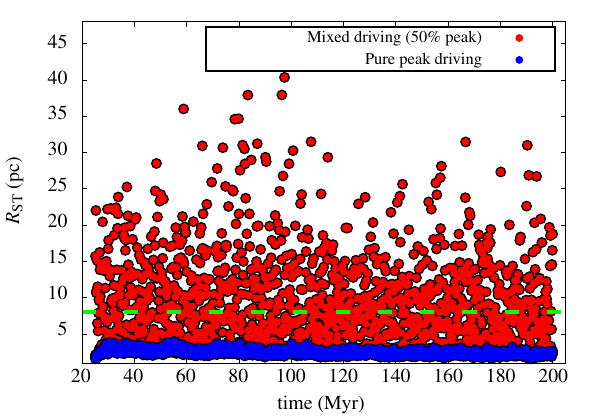}
\includegraphics[width=0.49\textwidth]{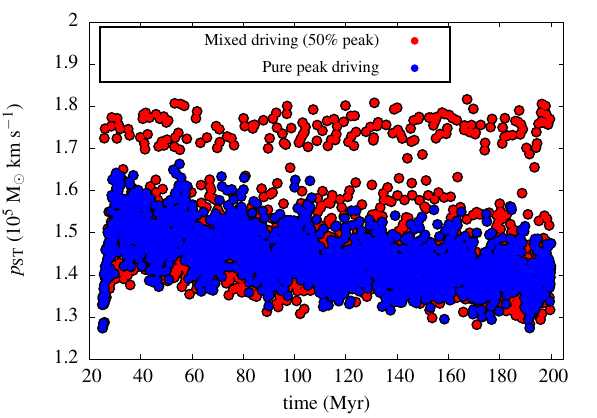}
\caption{Radius at the end of the Sedov-Taylor phase (eq. \ref{RST}; left) and injected momentum (eq. \ref{PST}; right) for runs with pure peak driving (run P-C-n$_3$) and mixed driving with $f_\mathrm{peak} =50$\% (run M50-C-n$_3$).
The dashed lines shows the resolution limit of $4\Delta x$, below which SNe are modelled via momentum input.}\label{fig:RST}
\end{figure*}
%%%%%%%%%%%%%%%%%%%%%%%%%%%%%%%%%%%%%%%%%%%%

\subsection{ Combined thermal energy and momentum injection}\label{sec:momentum}
%%%%%%%%%%%%%%%%%%%%%%%%%%%%%%%%%%%%%%%%%%%%
\begin{figure*}
\includegraphics[width=0.49\textwidth]{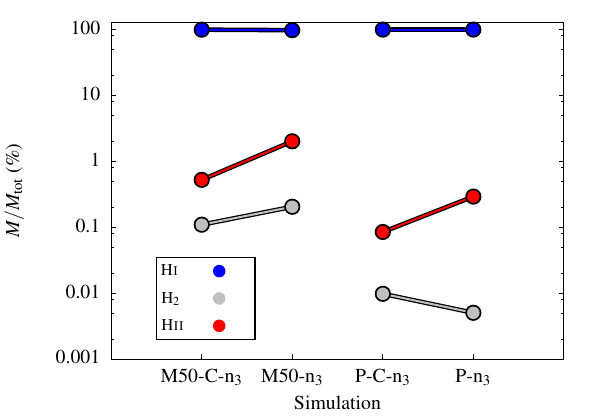}
\includegraphics[width=0.49\textwidth]{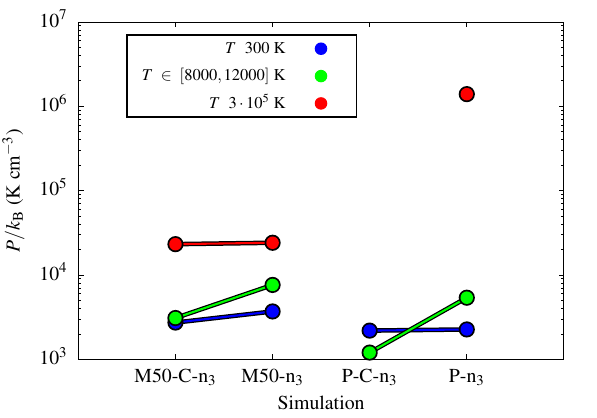}
\includegraphics[width=0.49\textwidth]{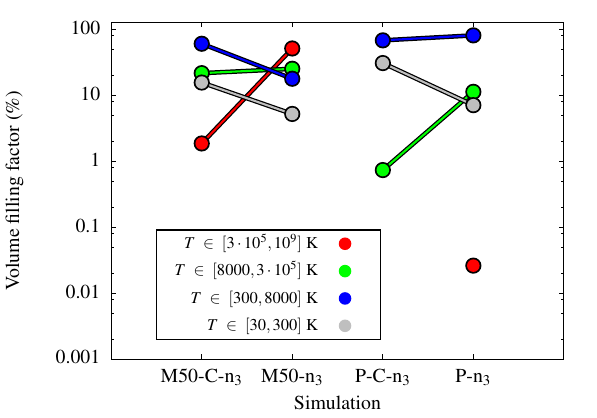}
\includegraphics[width=0.49\textwidth]{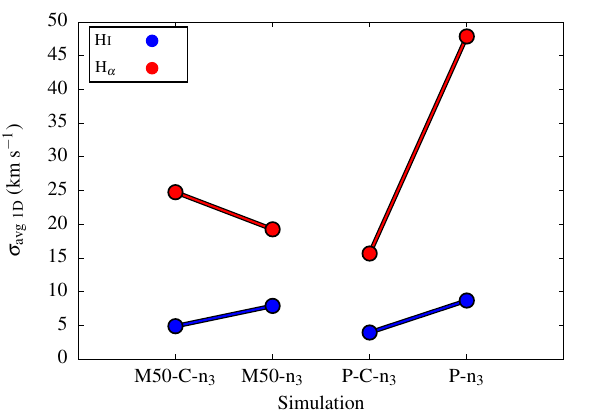}
\caption{Average mass fractions (top left), pressures (top right),VFFs (bottom left) and 1D velocity dispersions (bottom right) for the momentum input simulations with $n_\mathrm{i}=3 \cm$ and $\dot{N}_{\mathrm{SN,KS}}$. The points are: mixed driving with 50\% peak with combined thermal and momentum injection (M50-C-n$_3$); mixed driving with 50\% peak with thermal energy injection (M50-n$_3$); pure peak driving with combined injection (P-C-n$_3$); pure peak driving with thermal energy injection (P-n$_3$). The values are the average over the last 5 Myr of each simulation. The different lines connect simulations with same $n_\mathrm{i}$, SN rates, and driving modes, but different injection methods.}\label{fig:compare-last}
\end{figure*}
%%%%%%%%%%%%%%%%%%%%%%%%%%%%%%%%%%%%%%%%%%%%
Due to our finite numerical resolution, the injection of thermal energy alone might
lead to over-cooling in dense regions (see section \ref{sec:overcool} and Figure \ref{fig:TiSN}). Therefore, we introduce a SN model, which allows us to switch from
thermal energy to momentum input when the density in the vicinity of the SN is high
and the Sedov-Taylor phase is unresolved (see section \ref{combined_model}).
Using two simulations, we discuss how the combined model performs with respect to
the thermal energy injection scheme. In particular, we redo run P-n$_3$, i.e. peak
driving at $n_\mathrm{i}= 3 \cm$, and run M50-n$_3$, i.e. mixed driving with
$f_\mathrm{peak} = 50$\% and $n_\mathrm{i}= 3 \cm$. The corresponding new runs,
which use the combined model, are called P-C-n$_3$ and M50-C-n$_3$.

The comparison between runs with and without momentum input suffers from uncertainties in the resolution of peak SNe with thermal energy injection. If unresolved, these SNe are prone to cool too quickly and the resulting momentum input could be underestimated (see section \ref{combined_model} and \ref{sec:higres}). On the other hand, the momentum injection method only takes into account the momentum-generating Sedov-Taylor phase and neglects the additional contribution from the pressure-driven snowplough phase (section \ref{combined_model}).
Due to these intrinsic differences, a very detailed comparison is not possible at the moment.

In Figure~\ref{fig:RST} (left panel) we show that the Sedov-Taylor radius (eq.~\ref{RST}) is unresolved, i.e. $R_{\mathrm{ST}} < 4\Delta x$ as indicated
by the green dashed line, for all SNe in run P-C-$n_3$ and for 63\% of
the SNe in M50-C-$n_3$. In the latter case, the percentage of unresolved SNe being $>
50$ \% reflects the fact that most random SNe explode in a low density environment,
but not all of them. For the unresolved SNe, the injected momentum varies between
$1-2\times 10^5 \mo\ \mathrm{km\ s^{-1}}$, as shown in the right plot of Figure \ref{fig:RST}. On a side note, if we instead require that the temperature within the
injection region is higher than $10^6$ K, we find that $R_{\mathrm{ST}}$ as given in
eq. (\ref{RST}) is somewhat conservative and could safely be increased by a factor
of $\sim 1.6$.

In Figure~\ref{fig:compare-last} we compare the properties of the resulting ISM for runs with pure thermal energy injection and the new combined SN scheme. Here, the lines connect the models with the old and the new combined model. 

{\sc Chemical composition:} For both cases (peak and mixed driving), the two SN injection schemes give comparable mass fractions in all species. There are small differences, but these are well within the statistical fluctuations in the time evolution of each component (see Appendix for a discussion on the fluctuations). 

{\sc Gas pressure:} The main difference between the thermal energy injection and the combined SN injection scheme is that there is no hot gas present in case of the combined model, if all the SNe are using the momentum input scheme. This is the case for run P-C-n$_3$ (see also Fig.\ \ref{fig:RST}) and therefore we cannot compare the pressures of the hot phase between the two approaches in case of peak driving. If there is a random component (run M50-n$_3$), then the pressures of the hot phase are the same for thermal energy injection and combined model. The reason is that, in case of thermal energy injection, SNe which explode in dense gas cool on short time scales and do not contribute to the hot gas phase. 

Furthermore, both mixed and peak driving show a slightly reduced pressure of the warm phase when the combined scheme is applied. For the mixed driving case (run M50-n$_3$), this brings the warm and the cold phase into pressure equilibrium. For the peak driving case (run P-C-n$_3$), the pressure of the warm phase appears to be smaller than the pressure of the cold phase. This is an artefact. Since the young SN bubbles themselves contribute significantly to the warm phase (because we set the temperature within the injection region to $10^4$ K when momentum is injected), the warm phase is not formed self-consistently in this simulation.
However, the thermal energy injected in this way is between $1-9\ \%\ E_\mathrm{SN}$, with the majority of SNe lying in the range $1-5\ \%$, in very good agreement with \cite{KimOstriker14}, \cite{Martizzi+14} and \cite{WalchNaab14}.

{\sc Volume Filling Fractions:} The VFFs of the cold gas increase when using the momentum input scheme, 
which is more efficient in dispersing dense gas than the thermal injection method. As described above, the hot phase is missing or negligible when modelling SNe with momentum injection. 

{\sc H{\sc i} and H$_\alpha$ velocity dispersion:} The derived one dimensional $\HI$ velocity dispersion is slightly lower for the combined model. The H$_\alpha$ emission is sensitive to the actual physical state of the warm component. Setting an upper temperature value of $10^4$ K within the SN remnants strongly influences the derived H$_{\alpha}$ velocity dispersion. We do not recommend to trust the H$_{\alpha}$ velocity dispersion if the thermal state of the gas within the injection region has not been derived in a self-consistent way.
%%%%%%%%%%%%%%%%%%%%%%%%%%%%%%%%%%%%%%%%%%%%
\begin{figure*}
\includegraphics[width=0.49\textwidth]{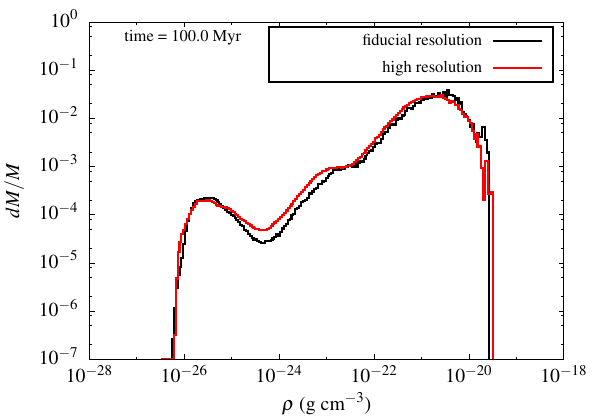}
\includegraphics[width=0.49\textwidth]{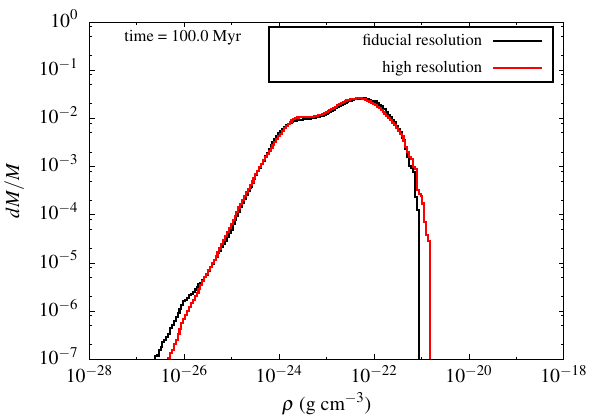}
\caption{Mass-weighted density PDFs for random (left column, runs R-n$_3$ and R-HR-n$_3$) and peak driving (right column, runs P-n$_3$ and P-HR-n$_3$) 
for simulations with $\Delta x = 2$~pc (fiducial resolution)
  or~1~pc (higher resolution), and our fiducial parameters $n_\mathrm{i}=3 \cm$ and $\dot{N}_{_{\rm SN, KS}}$.}\label{fig:respdf}
\end{figure*}
%%%%%%%%%%%%%%%%%%%%%%%%%%%%%%%%%%%%%%%%%%%%
\subsection{ Higher resolution}\label{sec:higres}
As an alternative one may avoid having $T_{\mathrm{i, SN}} < 10^6$ K in the SN injection region by going to a higher numerical resolution.
We run two additional simulations with resolution $\Delta x=1$ pc with pure peak and random driving (with pure thermal injection) for our
fiducial setup ($n_\mathrm{i}=3 \cm$ and $\dot{N}_{_{\rm SN, KS}}$).
For peak driving, this increase of a factor of two in resolution shifts the fraction of SNe having an estimated $T_{\mathrm{i, SN}} > 10^6$ K
from 65\% to 96\%, while for random driving this value is close to unity in both cases.
As shown in Figure~\ref{fig:hrappendix}, simulations with different resolutions do not display any substantial discrepancy in their temporal evolution. 

In Figure~\ref{fig:respdf} we compare the mass-weighted density PDFs for the fiducial and high resolution simulations for random (left panel) 
and peak driving (right panel).
We plot the mass-weighted PDF as this emphasises differences in the high density regime, which is most critical. The density PDFs also show 
good convergence between fiducial and higher resolution, with an increase in density of rather less than a factor of two.

\cite{KimOstriker14} show that the conditions $R_{\mathrm{ST}} > 3\ \Delta x$ and $R_{\mathrm{ST}} > 3\ R_{\mathrm{inj}}$ should be satisfied in order to recover a final momentum close to the values retrieved from their high resolution, sub-pc simulations.
Combining these criteria and having assumed $R_{\mathrm{inj}} \geqslant 4\ \Delta x$ (eq. \ref{Rinj}), their requirements can be translated into $R_{\mathrm{ST}} > 12\ \Delta x$. This slightly differs from our criterion, where $R_{\mathrm{ST}} \geqslant R_{\mathrm{inj}} > 3\ \Delta x$.
From eq. (\ref{RST}) (but see also section \ref{sec:momentum}) $R_{\mathrm{ST}} \approx 30.6\ n^{-0.4}$ pc. 
Therefore one can show that the critical density of the injection region (ambient medium) below which a SN is considered to be resolved is $n_\mathrm{crit}\approx 2 \cm$ from the \cite{KimOstriker14} criteria and $n_\mathrm{crit}\approx 59 \cm$ from ours at our fiducial resolution of 2 pc. For $\Delta x = 1$ pc, $n_\mathrm{crit}\approx 10 \cm$ from \cite{KimOstriker14} and $n_\mathrm{crit}\approx 330 \cm$ for us.
For an initial density of $3 \cm$, the highest densities reached are $n \sim 100-500 \cm$ (see Fig. \ref{fig:respdf}, right panel) and, hence, only a fraction of SNe in the high resolution run can be considered to be resolved. At fiducial resolution both ours and \cite{KimOstriker14} conditions reveal that all of the SNe are not resolved (see also left panel of Fig. \ref{fig:RST}). 
Therefore, simulations with peak driving and thermal injection with initial density $n_\mathrm{i} \geqslant 3 \cm$ suffer from an underestimate of the momentum input by SNe and hence should be considered representative of a lower limit of the impact of peak SNe.

\section{ Conclusions}\label{sec:conclusions}
{\sc Summary:} In this work, we perform 3D hydrodynamic simulations to study the SN-driven ISM in periodic volumes of size $(256\;{\rm pc})^3$. We include radiative cooling and diffuse heating, shielding from dust, and molecular gas (self-)shielding, as well as a chemical network to follow
H{\sc i}, H{\sc ii}, H$_2$, CO and C{\sc ii}. We study media with different mean gas densities,  $n_\mathrm{i} = 0.5,\;1,\;3,\;10,\;{\rm and}\;30 \;{\rm cm}^{-3}$, which, integrated over the box side, correspond to surface densities of $\Sigma_{\rm gas} \approx 4,\;8,\;24,\;81,\;{\rm and}\ 243 \;{\rm M}_\odot {\rm pc}^{-2}$. After an initial phase of turbulent stirring at root-mean-square velocity of $10 \kms$ for one crossing time (25 Myr), we switch to supernova (SN) driving and follow the simulations for more than three additional crossing times. 

The SN rates are constant for each simulation, but change with the mean box density according to the Kennicutt-Schmidt relation. We additionally perform models with a factor of two higher and lower SN rates. We explore the impact of the placement of the SNe relative to the dense gas. We distinguish between random positions (random driving), positioning on local density peaks (peak driving), and a mixture of the two (mixed driving). 
The different SN placements should reflect that SNe could explode (i) mostly in their birthplaces, where they might be deeply embedded (peak driving); (ii) preferentially in low density environments (random driving). Random driving provides a simple way to introduce SNe in evolved, low density environments without including all the necessary extra physics (wind and ionisation feedback from massive stars; clustering of massive stars; runaway stars) acting in between the point of star formation and the SN explosion.
For mixed driving, we perform a set of simulations where we explore different ratios of peak to random driving using the fiducial setup ($n_\mathrm{i} = 3\;{\rm cm}^{-3}$ and $\dot{N}_\mathrm{SN, KS}$).\\

{\sc Random driving at different supernova rates:} In the case of random driving, most of the mass is in cold, dense atomic and molecular hydrogen, whereas most of the volume is filled with hot, rarefied gas. The ISM is out of pressure equilibrium in all simulations apart from two cases with lowest density and lowest SN rate. The synthetically observed 1D velocity dispersions are $\sim 5-7 $ $\kms$ in H{\sc i} and $\sim 17 - 25 $ $\kms$ in H$_\alpha$. 
Increasing the SN rate by a factor of two leads to an increase in the mean gas pressure of the different phases (cold, warm, hot medium), a small increase in the velocity dispersion ($\sim$ few $\kms$), and higher H{\sc i} mass fractions, while the H$_2$ mass fractions decrease. However, the volume is always completely filled with hot gas, unless we consider low densities ($n_\mathrm{i}\le1\;{\rm cm}^{-3}$) and SN rates of $\dot{N}_\mathrm{SN} \lesssim 2$ Myr$^{-1}$. We attribute this behaviour to a thermal runaway process (see section \ref{sec:thermal}), which occurs at high pressures, where only the cold branch of the equilibrium cooling curve can be reached, and no two-phase medium can form. In our case, though, the pressures are artificially determined. The boxes are not allowed to reach hydrostatic equilibrium to a larger galactic potential, so the SNe continuously feed energy into the high-pressure, hot medium. Thus, they can push almost all of the gas mass into small, molecular clumps that quickly 
form 
molecular hydrogen.\\

{\sc Peak and mixed driving:} 
For peak instead of random driving, the ISM has a completely different structure. It is dominated by a filamentary distribution of warm gas (mostly H{\sc i}), with little to no hot gas present. SNe that explode in dense gas also disperse the cold medium, and therefore the mass fraction of molecular hydrogen is small.
The absence of hot gas (volume filling fraction VFF $\ll$ 50\%) is due to the low heating efficiency and strong cooling of the SNe, which interact with dense gas. However, due to our limited resolution, we note that the impact of peak SNe is probably underestimated in high density regions (for $n > 1-60 \cm$ with fiducial and $n > 10-300 \cm$ with high resolution).
In these cases, one should anyway expand the model to include other important physical conditions and processes, such as stellar clustering, stellar winds, and ionising radiation, and non-constant SN rate are critical ingredients, that shape the ISM and have to be taken into account.

For the setup with $n_\mathrm{i}=3\;{\rm cm}^{-3}$ and $\dot{N}_\mathrm{SN, KS}$, we vary the fraction of peak driving to random driving to explore the effect of a mixed SN placement at different fractions $f_\mathrm{peak}$. A relatively sharp transition between the two regimes (peak and random) occurs when $\sim$50\% of the SN are located within density peaks. As $f_\mathrm{peak}$ increases, the mass fraction in H$_2$ drops, and the VFF of the warm and cold gas decreases. Interestingly, we find that the pressure of the hot phase as well as the velocity dispersion in H$_\alpha$ increase with increasing peak fraction (H$_\alpha \sim 50$ $\kms$ for 100\% peak driving). This behaviour can be attributed to the younger age of the SN remnants that contribute to these quantities.\\

{\sc Combined energy and momentum input:} 
In low density gas, the SNe are well resolved and we model them with thermal energy input. Explosions within high density regions are eventually unresolved and would be subject to strong radiative (over-)cooling. Therefore, we introduce a new model, which combines thermal energy input for resolved SNe and momentum input for unresolved SNe. The momentum input at the end of the Sedov-Taylor phase is calculated using the relations derived in \citet{Blondin+98}. We put the model to work in two of the simulations (peak driving and mixed driving at $f_\mathrm{peak}=50$\%). We find that the momentum input model fails to produce any hot gas because the shock speeds are too small to heat the medium to more than $10^4$ K. Otherwise, the combined model gives similar results to the thermal energy injection model and is therefore a viable alternative to model SN in partly unresolved environments -- with the limitation that the temperature structure of the gas is no longer self-consistent.\\

{\sc Clumpy H$_2$ in gas-rich discs:} When thermal runaway sets in, hot gas at high pressure pushes the gas into small and dense clouds, leading to a
fast and efficient conversion from \HI\ to H$_2$. Warm gas directly cools towards the cold branch of the equilibrium curve and, similar to the case of a high far-UV interstellar radiation field, a bi-stable equilibrium between the cold and warm phases does not exist anymore. We speculate that the molecular-dominated, very clumpy structure of the ISM in simulations with a high gas surface density $\Sigma_\mathrm{gas}\gtrsim 100\;{\rm M}_{\odot}\;{\rm pc}^{-2}$ could be a reasonable representation of systems with high gas surface densities and very high mid-plane pressures like ULIRGs \cite[][but see also \citealt{BlitzRosolowsky05}]{DownesSolomon98} or normal star forming galaxies at high redshift \citep{Genzel+10} that have high gas fractions and SFRs \citep[e.g.][]{Tacconi+10,Tacconi+13}. 
Their mid-plane pressures are three to four orders of magnitudes higher than for the Milky Way \citep{Bowyer+95,Berghoefer+98,JenkinsTripp11} and
plausibly reach values of $P/k_\mathrm{B} \sim 10^{6-7}$ K$\cm$ for surface densities of 100 M$_\odot$ pc$^{-2}$ and above \citep{Swinbank+11}, very similar to our models for $n_\mathrm{i}=10$ or~$30 \cm$ with random driving. Such high mid-plane pressure forces all the gas onto the cold branch of the equilibrium curve, so that it reaches densities that allow quick conversion from atomic to molecular gas. Thus, in high pressure, gas rich environments, the sizes and masses of the collapsing and H$_2$ forming structures could be regulated by supernova feedback. Our simulations show that the mass budget of the ISM can be dominated by molecular gas while, at the same time, this molecular gas is still found in small dense clumps with low filling factor surrounded by hot rarefied gas, rather than being evenly distributed. However, this picture needs to be refined with simulations of stratified, high-surface density discs \citep[similar to][]{ShettyOstriker12} and a more self-consistent treatment of star formation 
to 
address the issue of self-regulation in gas-rich environments.  
On the other hand, for low $\Sigma_\mathrm{gas}\sim 5\;{\rm M}_{\odot}\;{\rm pc}^{-2}$, where the ISM has reasonable pressures and VFFs, we find too little molecular gas. Here we are probably missing the aid of self-gravity.\\

{\sc Limitations of the model:} For peak driving, the limited resolution employed raises concerns about the effectiveness of density peak SNe. Recent results from \cite{KimOstriker14} suggest that our models with pure thermal injection should be considered to be lower limits on the impact of peak driving on the ISM. For random SNe, in particular for intermediate $\Sigma_\mathrm{gas}\sim 10-100\;{\rm M}_{\odot}\;{\rm pc}^{-2}$, the creation of high pressure, high VFF, hot gas in the random driving case is overestimated due to our choice of periodic boundary conditions in all three dimensions. In a
stratified disc, these high pressures would power a galactic fountain or outflow, lowering the ambient pressure in the mid-plane to the hydrostatic
equilibrium value within a crossing time (few dozens of Myr) of a scale height for the hot gas. Other key physical ingredients, such as ionising radiation, stellar winds, and stellar clustering, etc. could also play a crucial role in determining the state of the multi-phase ISM. For these reasons, the employment of the simplified setup presented here does not allow us to draw any quantitative conclusions. However, this kind of study clearly shows the qualitative consequences of each model.
%%%%%%%%%%%%%%%%%%%%%%%%%%%%%%%%%%%%%%%%%%%%
%%%%%%%%%%%%%%%%%%%%%%%%%%%%%%%%%%%%%%%%%%%%

\section{Acknowledgements}
We thank an anonymous referee for comments and suggestions that improved the clarity of this work. We also thank A. Pardi and A. Bubel for pointing out some minor errors in an earlier version of the paper.
We thank A. Ballone and A. Cald{\'u}-Primo for stimulating discussions.
The authors acknowledge the Deutsche Forschungsgemeinschaft (DFG) for funding
through the SPP 1573 ``The Physics of the Interstellar Medium''. S.~Walch acknowledges funding by the Bonn-Cologne-Graduate School.
M-M.~Mac~Low acknowledges support from NSF grant AST11-09395 and the Alexander von Humboldt-Stiftung.
R. S. Klessen and S. Glover acknowledge support from the DFG via SFB 881 ``The Milky Way System'' (sub-projects B1, B2 and B8).
R. S. Klessen furthermore acknowledges support from the European Research Council under the
European Community’s Seventh Framework Programme (FP7/2007-2013) via the ERC
Advanced Grant STARLIGHT (project number 339177).
R.~W\"unsch acknowledges support by project P209/12/1795 of the Czech Science Foundation and by project RVO:~67985815.
T.~Peters acknowledges financial support through a Forschungskredit of the University of Z\"{u}rich, grant no. FK-13-112. A. Gatto visited the AMNH with support from a Kade Fellowship during completion of this work.
The software used in this work was in part developed by the DOE NNSA-ASC OASCR Flash Center at the University of Chicago.
We thank C. Karch for the program package \textsc{FY} and M. Turk and the \textsc{yt} community for the \textsc{yt} project \citep{yt}.
The authors gratefully acknowledge the Gauss Centre for Supercomputing
for funding this project by providing support and high-performance computing time on SuperMUC at
the Leibniz Supercomputing Centre (LRZ). Additional simulations have been performed on the Odin and Hydra
clusters at the Rechenzentrum Garching (RZG).
%%%%%%%%%%%%%%%%%%%%%%%%%%%%%%%%%%%%%%%%%%%%

\bibliographystyle{mn2e}
\bibliography{Modelling_the_supernova-driven_ISM_in_different_environments}
%%%%%%%%%%%%%%%%%%%%%%%%%%%%%%%%%%%%%%%%%%%%
%%%%%%%%%%%%%%%%%%%%%%%%%%%%%%%%%%%%%%%%%%%%

\appendix
\section{Temporal evolution vs. time average}\label{sec:appendix}
In this section we show the differences between the analysed
quantities averaged over the last 5 Myr of each simulation and their
full temporal evolution.
Generally, due to our choice of stopping the simulations once a chemodynamical equilibrium is reached, different runs
have different final times. This already introduces uncertainty when comparing different simulations with quantities averaged over the last 5 Myr.

\begin{figure*}
\includegraphics[width=0.45\textwidth]{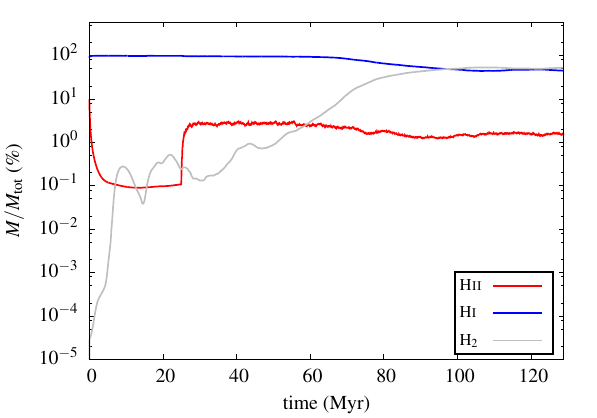}
\includegraphics[width=0.45\textwidth]{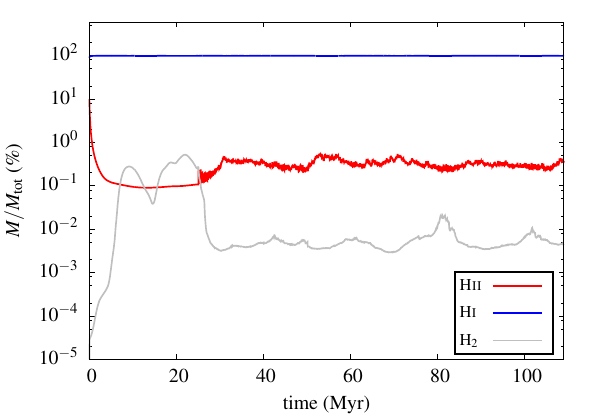}
\includegraphics[width=0.45\textwidth]{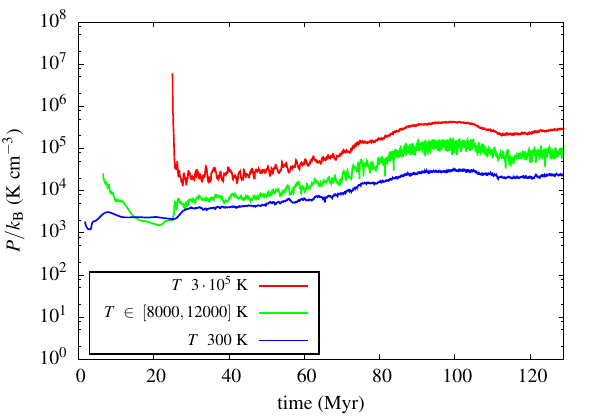}
\includegraphics[width=0.45\textwidth]{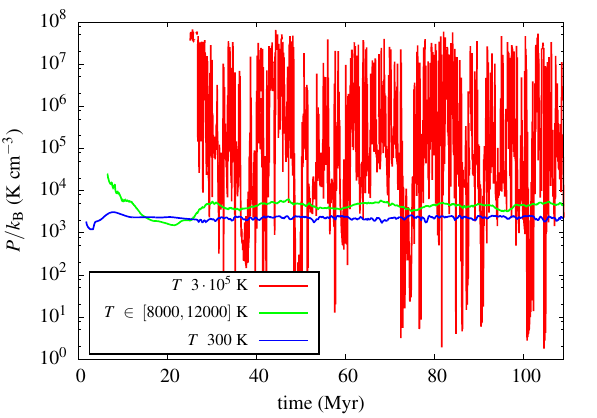}
\includegraphics[width=0.45\textwidth]{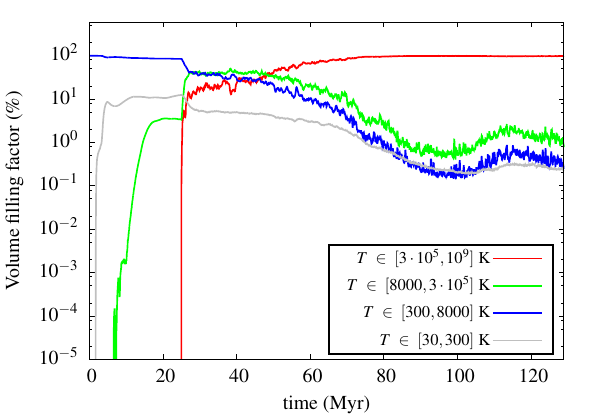}
\includegraphics[width=0.45\textwidth]{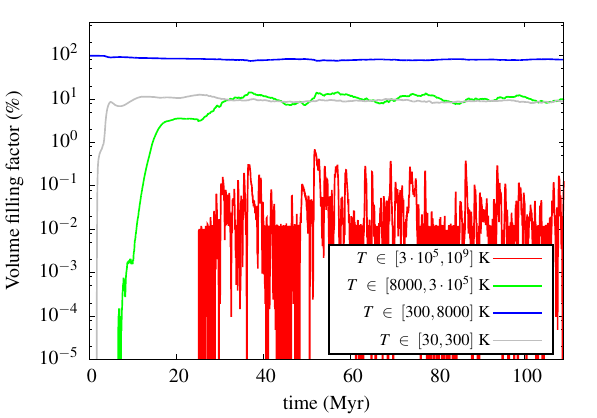}
\includegraphics[width=0.45\textwidth]{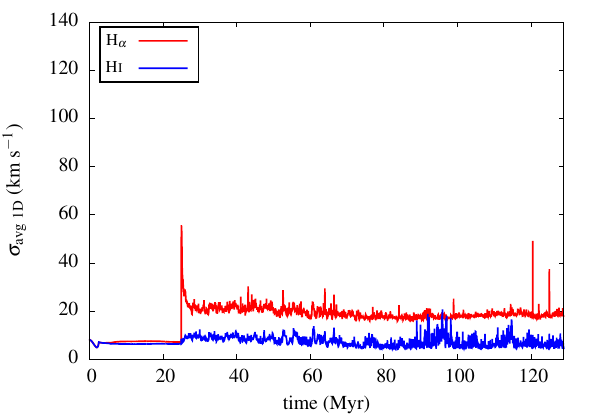}
\includegraphics[width=0.45\textwidth]{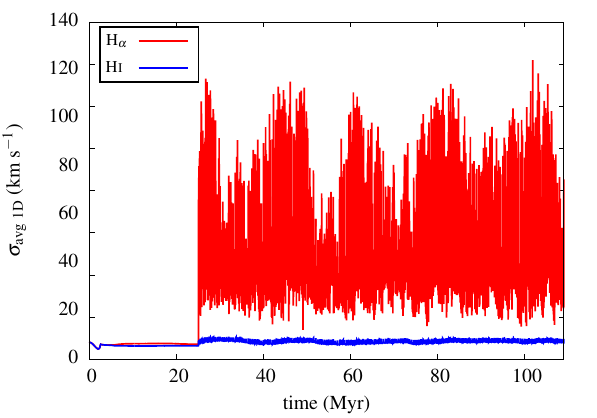}
\caption{Evolution of mass fractions, pressures, VFFs and velocity dispersions for $n_\mathrm{i}=3
  \cm$ and $\dot{N}_{_{\rm SN, KS}}$ with random  
(left column) and peak driving (right column).}\label{fig:time-comparison}
\end{figure*}

\begin{figure*}
\Large
\includegraphics[width=0.45\textwidth]{plots/app-p-m.pdf}
\includegraphics[width=0.45\textwidth]{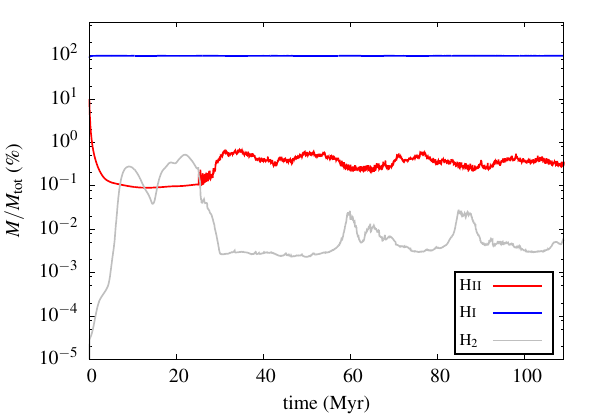}
\includegraphics[width=0.45\textwidth]{plots/app-p-pt.pdf}
\includegraphics[width=0.45\textwidth]{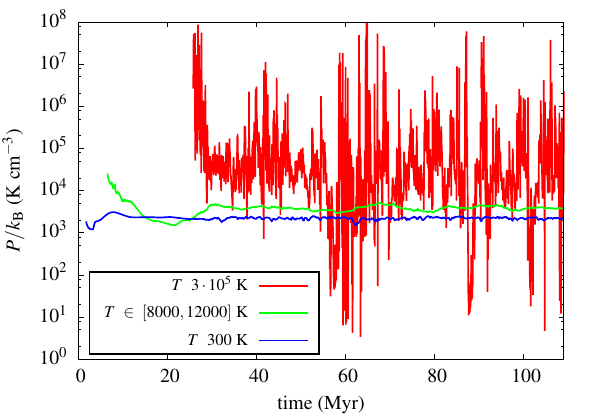}
\includegraphics[width=0.45\textwidth]{plots/app-p-fv.pdf}
\includegraphics[width=0.45\textwidth]{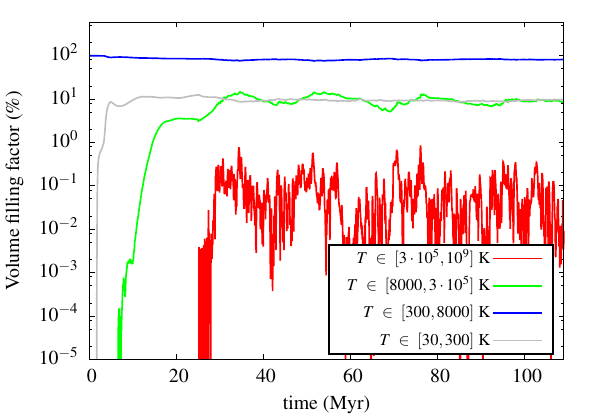}
\includegraphics[width=0.45\textwidth]{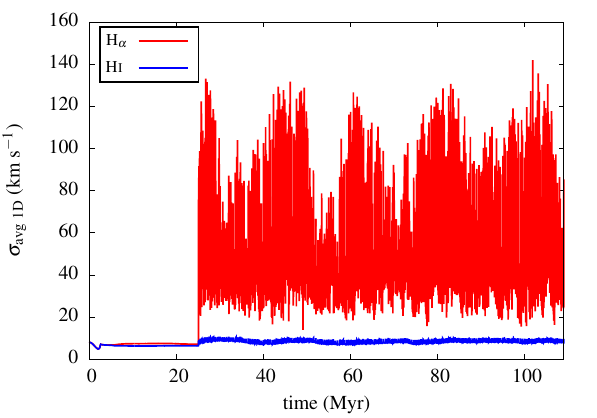}
\includegraphics[width=0.45\textwidth]{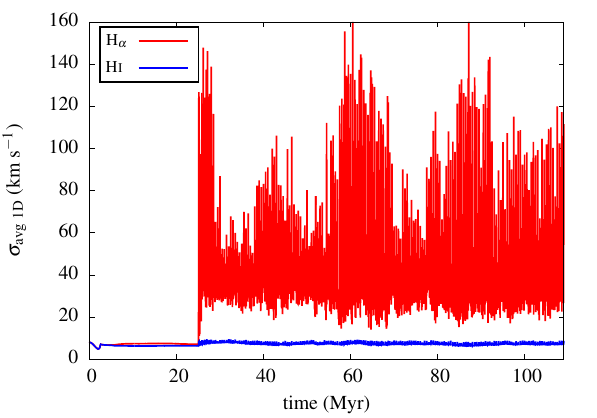}
\caption{Evolution of mass fractions, pressures, VFFs and velocity dispersions for $n_\mathrm{i}=3
  \cm$ and $\dot{N}_{_{\rm SN, KS}}$ with peak driving 
for the fiducial resolution of $\Delta x = 2$ pc (left column) and for higher resolution of $\Delta x = 1$ pc (right column).}\label{fig:hrappendix}
\end{figure*}

Figure~\ref{fig:time-comparison} shows the evolution of mass fractions, VFFs, pressures, and velocity dispersions for the fiducial setup with
pure random (left column) and pure peak driving (right column).
For random driving the evolution of the quantities is smooth and the time average fairly portrays the global trend.
On the other hand, for peak driving this is not true anymore. Highly
time-varying H$_2$ mass fraction, $\sHa$, hot gas pressure, and
VFF cause the 
time average to be not completely representative of the temporal trend.
This behaviour is also important when comparing simulations with same driving mode, as shown in Figure~\ref{fig:hrappendix}
for peak driving with fiducial (left column) and high resolution (right column).
This effect, although not  significant, should be taken
  into account when comparing simulations.
%%%%%%%%%%%%%%%%%%%%%%%%%%%%%%%%%%%%%%%%%%%%
%%%%%%%%%%%%%%%%%%%%%%%%%%%%%%%%%%%%%%%%%%%%
%%%%%%%%%%%%%%%%%%%%%%%%%%%%%%%%%%%%%%%%%%%%

\label{lastpage}

\end{document}